\documentclass[iop]{emulateapj}
\usepackage{amsmath}
\usepackage{subfigure}

\newcommand{\E}[1]{\ensuremath{\times10^{#1}}}
\newcommand{\low}{{\ell}\text{ow}}
\newcommand{\hHz}{\ensuremath{\text{hHz}}}
\newcommand{\LF}{\ensuremath{\text{LF}}}
\newcommand{\cut}{\ensuremath{\text{cut}}}

\slugcomment{} 

\shorttitle{The X-ray variability of SAX~J1808.4--3658}
\shortauthors{Bult and van~der~Klis}

\begin{document}

\title{The aperiodic X-ray variability of the accreting millisecond pulsar SAX~J1808.4--3658}

\author{Peter Bult and Michiel van der Klis}
\affil{	
	Anton Pannekoek Institute, University of Amsterdam,
	Postbus 94249, 1090 GE Amsterdam,  The Netherlands
}

\begin{abstract}
    We have studied the aperiodic variability of the 401~Hz accreting millisecond
    X-ray pulsar SAX~J1808.4--3658 using the complete data set collected with
    the {\it Rossi X-ray Timing Explorer} over 14 years of observation. 
    The source shows a number of exceptional aperiodic timing phenomena that are observed against a
    backdrop of timing properties that show consistent trends in all five observed 
    outbursts and closely resemble those of other atoll sources.
    We performed a detailed study of the enigmatic $\sim$410~Hz QPO, which has only been
    observed in SAX~J1808.4--3658. We find that it appears only when the upper kHz QPO frequency is 
    less than the 401~Hz spin frequency. The difference between the $\sim$410~Hz QPO frequency and the
    spin frequency follows a similar frequency correlation as the low frequency
    power spectral components, suggesting that the $\sim$410~Hz QPO is a
    retrograde beat against the spin frequency of a rotational phenomenon in the 9~Hz
    range.
    Comparing this 9~Hz beat feature with the Low-Frequency QPO in SAX~J1808.4--3658 and
    other neutron star sources, we conclude that these two might be part of the same 
    basic phenomenon. We suggest that they might be caused by retrograde precession
    due to a combination of relativistic, classical and magnetic torques.
    Additionally we present two new measurements of the lower kHz QPO, allowing us,
    for the first time, to measure the frequency evolution of the twin kHz QPOs in this
    source. The twin kHz QPOs are seen to move together over 150~Hz, maintaining 
    a centroid frequency separation of $(0.446\pm0.009)\nu_{\text{spin}}$.
\end{abstract}


\section{Introduction}
    The transient X-ray source SAX~J1808.4--3658 (SAX~J1808) was discovered
    with the \textit{BeppoSax} satellite in September 1996 \citep{Zand1998}. It
    remained in quiescence for 1.6~yr after the discovery, until the
    \textit{Rossi X-ray Timing Explorer} (\textit{RXTE}) detected renewed X-ray activity
    with the Proportional Counter Array \citep{IAUC.6876}. The initial
    \textit{RXTE} observations revealed coherent pulsations at 401~Hz, making SAX~J1808
    the first known accreting millisecond X-ray pulsar (AMXP)
    \citep{Wijnands1998a}.
    
	From extrapolating the recurrence time between the 1996 and
    1998 outbursts, SAX~J1808 was expected to show a new cycle of activity
    around November 1999. Unfortunately, Solar constraints prevented
    observations until late January 2000, at which time SAX~J1808 was observed
    to be in a low luminosity state at roughly a tenth of the typical flux of
    the 1998 outburst \citep{Klis2000}. It is generally assumed that the main
    outburst occurred when the Solar constraints prevented observation \citep{Wijnands2001}.
    The low luminosity state persisted for $\sim114$~days \citep{Wijnands2001}, 
    showing dramatic changes in luminosity. On a timescale of $\sim$5~days the 
    luminosity would drop below a detection limit of $\sim10^{32}$~erg~s$^{-1}$ 
    and then increase again to $\sim10^{35}$~erg~s$^{-1}$.
	During these episodes of increased emission the 401~Hz pulsations were detected 
	along with a sporadic strong
    modulation at about 1~Hz \citep{Klis2000}. \citet{Campana2008} proposed
    that the 5-day episodic variations of luminosity were caused by an
    intermittently active propellor effect \citep{Illarionov1975}, periodically
    inhibiting accretion onto the neutron star. The 1~Hz flaring remained
    unique to SAX~J1808 until a similar feature was discovered in the AMXP NGC
    6440 X-2 \citep{Atel.2672, Patruno2013}.
	
    The fourth outburst of SAX~J1808 was detected in October 2002, some 2.8~yr
    after the 2000 outburst, triggering an extensive monitoring campaign
    with \textit{RXTE}. Twin kHz quasi-periodic oscillations (QPOs) were detected early in the outburst
    \citep{Wijnands2003}, for the first time in a source for which the spin
    frequency was precisely known. An additional QPO was unexpectedly detected
    at 410 Hz, which so far is unique to SAX~J1808 and whose nature remains a
    mystery. After the main outburst, which lasted several weeks, the source
    entered a low luminosity state showing the same $\sim$5~day episodes of
    increased emission and 1~Hz flaring seen in 2000 \citep{Wijnands2004}.  

    Since its 2002 outburst, SAX~J1808 has been observed in outburst another three
    times; in 2005, 2008 and 2011. 
    For both the 2005 and 2008 outbursts a low-luminosity prolonged outburst tail was
    observed, again showing luminosity variations on a $\sim5$~day timescale. During 
    these episodes the 1~Hz flaring 
    was again observed in 2005 but not in 2008 \citep{Patruno2009}. 
    Observations of the 2011 outburst were cut short by
    Solar constraints, such that only the onset of the low-luminosity outburst tail was seen,
    where again, no 1~Hz flaring was detected \citep{Patruno2012}. The 2008 and
    2011 outbursts were later found to show a 1--5~Hz flaring which shares many
    of the characteristics of the 1~Hz flaring, but occurred at luminosities an
    order of magnitude higher, near the peak of the outburst
    \citep{Bult2014}.  

    SAX~J1808 showed several thermonuclear X-ray bursts in each outburst.
    Using the X-ray bursts of the 1998, 2002 and 2005 outbursts
    \citet{Galloway2006} derived a distance estimate of $3.5\pm0.1$~kpc.

    The literature contains a large body of work on the coherent pulsations of
    SAX~J1808 \citep{Wijnands1998a, Poutanen2003, Burderi2006, Hartman2008, Leahy2008, 
    Hartman2009, Ibragimov2009, Patruno2012}. The study of the stochastic variability 
    of SAX~J1808, on the other hand, remains incomplete, with all work constrained to the earlier
    outbursts. Specifically, \citet{Wijnands1998b} studied the broadband power
    spectrum of the 1998 outburst and \citet{Menna2003} considered the coupling between
    the broad band noise and the coherent pulsations. The 2002 outburst power spectrum was studied
    by \citet{Wijnands2003}, but only the high frequency region, in the
    context of the kHz QPOs. A more general analysis of the 1998 and 2002
    outbursts was done by \citet{Straaten2005}, who compared the timing
    behavior of AMXPs with non-pulsating low-mass X-ray binaries. The stochastic variability
    of the 2005, 2008 and 2011 outbursts remains unaddressed. 
    
    With the demise of \textit{RXTE}, its 14~year monitoring campaign of SAX~J1808 has
    now been completed. We use this remarkable dataset to provide a complete
    overview of the stochastic time variability of SAX~J1808. In this work we
    analyze all outbursts in a consistent approach, but do not consider the low
    luminosity outburst tail, which has been studied in-depth by \citet{Patruno2009}.
    We provide the first in-depth analysis of the broadband power spectra of the 2005, 2008 and 2011
    outbursts and characterize the rich variability phenomenology observed in
    this accreting millisecond X-ray pulsar.

\section{Data Reduction}
    We used all \textit{RXTE} pointed observations of SAX~J1808's outbursts (see
    Table~\ref{tab:intervals} for ObsIDs), selecting only observations with
    stable pointing, source elevation above 10$^\circ$ and pointing offset less
    than 0.02$^\circ$.
    
\begin{deluxetable*}{llllcccc}
    \tabletypesize{\scriptsize}
    \tablecaption{
    	Interval Listing
        \label{tab:intervals} 
    }
    \tablewidth{ 1.0 \linewidth }
    \tablehead{
        \colhead{Interval} & \colhead{State} & \colhead{Start MJD} & \colhead{ObsIDs} 
        & \colhead{} & \colhead{}  & \colhead{} & \colhead{} 
    }
    \startdata
        \cutinhead{1998 Outburst}
            1  &  IS   & 50914.9 & A-01-03S			        	\\
            2  &  EIS  & 50919.8 & A-03-00 	    	        	\\
            3  &  EIS  & 50920.1 & A-04-00    		        	\\
            4  &  EIS  & 50921.3 & A-05-00		            	\\
            5  &  EIS  & 50921.5 & A-06-01		            	\\
            6  &  EIS  & 50921.7 & A-06-000		            	\\
            7  &  EIS  & 50921.9 & A-06-00		            	\\
            8  &  EIS  & 50923.9 & A-07-00	    	        	\\
            9  &  EIS  & 50926.8 & A-08-00		           		\\
           10  &  EIS  & 50927.7 & A-09-01, A-09-02           	\\
           11  &  EIS  & 50928.6 & A-09-03, A-09-04           	\\
           12  &  EIS  & 50929.8 & A-09-00		            	\\
           13  &  EIS  & 50930.6 & A-10-02, A-10-01, A-10-00	\\
        \cutinhead{2002 Outburst}
            1  &  IS   & 52562.1 & B-03-04-00							\\
            2  &  LLB  & 52562.3 & B-01-01-000						\\
            3  &  LLB  & 52563.1 & B-01-01-03, B-01-01-04			\\
            4  &  LLB  & 52563.4 & B-01-01-01 					            \\
            5  &  LLB  & 52564.3 & B-01-01-020, B-01-01-02		            \\
            6  &  IS  & 52565.0 & B-01-02-01, B-01-02-000, B-01-02-00				\\
            7  &  IS  & 52566.0 & B-01-02-02, B-01-02-03, B-01-02-04					\\
            8  &  IS  & 52567.1 & B-01-02-05, B-01-02-06, B-01-02-07					\\
            9  &  IS  & 52568.1 & B-01-02-10, B-01-02-08		            			\\
           10  &  IS  & 52569.0 & B-01-02-20					            			\\
           11  &  IS  & 52569.2 & B-01-02-09					            			\\
           12  &  IS  & 52569.4 & B-01-02-19, B-01-02-23		        	    			\\
           13  &  IS  & 52570.0 & B-01-02-21, B-01-02-11		            				\\
           14  &  IS  & 52570.4 & B-01-02-18, B-01-02-12, B-01-02-22							\\
           15  &  IS  & 52571.0 & B-01-02-13, B-01-02-15, B-01-02-14, B-01-02-16					\\
           16  &  IS  & 52572.0 & B-01-02-17, B-01-03-00, B-01-03-04, B-01-03-05						\\
           17  &  IS  & 52572.2 & B-01-03-000																\\
           18  &  IS  & 52572.9 & B-01-03-06, B-01-03-07, B-01-03-010, B-01-03-01							\\
           19  &  IS  & 52573.4 & B-01-03-08, B-01-03-09, B-01-03-10, B-01-03-11, B-01-03-12, B-01-03-13 	\\
           20  &  IS  & 52574.1 & B-01-03-02																	\\
           21  &  IS  & 52574.8 & B-01-03-14, B-01-03-03, B-03-05-00, B-02-01-04							\\
           22  &  IS  & 52577.1 & B-02-01-000, B-02-01-00											\\
        \cutinhead{2005 Outburst}
            1  &  IS   & 53523.0 & C-01-01													\\
            2  &  IS   & 53523.5 & C-01-02, C-01-03, C-01-04, C-01-05, C-02-00			\\
            3  &  IS   & 53524.9 & C-02-01											\\
            4  &  IS   & 53525.8 & C-02-02										\\
            5  &  IS   & 53526.8 & C-02-030										\\
            6  &  IS   & 53527.0 & C-02-03										\\
            7  &  LLB  & 53527.8 & C-02-04, C-02-05, C-02-06						\\
            8  &  IS   & 53530.8 & C-02-08										\\
            9  &  IS   & 53531.3 & C-03-03, C-03-02, C-03-00, C-03-01			\\
           10  &  IS   & 53533.6 & C-03-05, C-03-04, C-03-07, C-03-08	        \\
           11  &  IS   & 53535.7 & C-03-06, C-03-09, C-03-11, C-03-10 	        \\
        \cutinhead{2008 Outburst}
            1  &  IS    & 54731.9 & D-01-01, D-01-00, D-01-02		            		\\
            2  &  IS    & 54732.7 & D-01-080, D-01-08, D-01-03				    			\\
            3  &  IS    & 54733.7 & D-01-04, D-01-07, D-01-06, D-01-05, D-01-10, D-02-00, D-02-01	\\
            4  &  IS    & 54736.5 & D-02-05, D-02-06, D-02-03, D-02-04		    			\\
            5  &  IS/F  & 54739.4 & D-02-07, D-02-09, D-02-02, D-02-08 				\\
            6  &  IS    & 54742.5 & D-03-00, D-03-01, D-03-08, D-03-02		    	\\
            7  &  IS    & 54745.5 & D-03-03, D-03-10, D-03-05, D-03-04		    	\\
        \cutinhead{2011 Outburst}
            1  &  IS/F  & 55869.9 & E-01-01, E-01-00								\\
            2  &  IS/F  & 55871.1 & E-01-02											\\
            3  &  IS/F  & 55872.0 & E-01-03, E-01-04								\\
            4  &  LLB/F & 55872.9 & E-01-05, E-01-06								\\
            5  &  LLB   & 55874.0 & E-01-070, E-01-07			        			\\
            6  &  IS    & 55875.2 & E-01-08, E-01-09, E-02-00 			    	    \\
            7  &  IS    & 55877.0 & E-02-01, E-02-02					    	   	\\
            8  &  IS    & 55878.0 & E-02-03, E-02-04						        \\
            9  &  EIS   & 55879.0 & E-02-05, E-02-06, E-02-07, E-02-08	 			\\
           10  &  EIS   & 55881.3 & E-02-09, E-02-10, E-03-00, E-03-01, E-03-04 	\\
    \enddata
\tablecomments{
\textit{RXTE} ObsIDs grouped together in intervals for all outbursts. ObsIDs are chronologically ordered from left to right and top to bottom.
Source states are abbreviated as: EIS - extreme island state; IS - island state; LLB - lower-left banana, F - low luminosity flaring present. 
A = `30411-01'; B = `70080', C = `91056-01', D = `93027-01', E = `96027-01'.\\
}
\end{deluxetable*}

    Using the Standard-2 data we created Crab normalized 2--16~keV light
    curves.  We also calculated the Crab normalized soft color as the count
    rate in the 3.5--6.0~keV band divided by the count rate in the 2.0--3.5~keV
    band. Similarly, the Crab normalized hard color is calculated by  dividing
    the 9.7--16~keV band count rate by the count rate in the 6.0--9.7~keV band
    (see e.g. \citealt{Straaten2003}, for the detailed procedure).

    The timing analysis was done using high time resolution (122~$\mu$s)
    Event and GoodXenon data, selecting all events in the 2--20~keV energy
    band.  Data obtained in GoodXenon mode were binned to 122~$\mu$s prior
    to further analysis.  Fourier transforms were calculated using
    256~s segments at full time resolution, yielding power spectra with a
    frequency resolution of $\sim$0.004~Hz and a Nyquist frequency of 4096~Hz.
    No background subtraction or dead-time correction was applied prior to
    the Fourier transform. The power spectra were calculated using the standard
    Leahy normalization \citep{Leahy1983}. Following the method of
    \citet{KleinWolt2004} we inspected the high frequency region ($> 1500$~Hz)
    for anomalous features. Since none were found we subtracted a Poisson noise
    power spectrum calculated using the analytical formula of \citet{Zhang1995} and shifted
    to fit the $> 1500$~Hz power. All power spectra were averaged per ObsID. If necessary, multiple consecutive
    ObsIDs were averaged into \textit{intervals} to improve statistics, but
    only if the power spectra showed similar morphology and the corresponding
    colors and intensities were consistent with being the same.  Details of the
    intervals used are given in Table~\ref{tab:intervals}.  Finally we
    estimated the background count rate using the \textsc{ftool pcabackest} and
    used it to renormalize the power spectra to units of source fractional rms
    squared per Hz \citep{Klis1995}. 

    The power spectra are described in terms of a sum of Lorentzian profiles
    (see e.g. \citealt{Nowak2000, Belloni2002}). Each profile, $L(\nu|\nu_{\max},Q,r)$, is 
    a function of Fourier frequency $\nu$ and determined by three
    parameters: the characteristic frequency $\nu_{\max} =
    \sqrt{\nu_{0}^2 + (\mbox{FWHM}/2)^{2} }$, the quality factor $Q =
    \nu_{0}/\mbox{FWHM}$ and its fractional rms amplitude squared
    $r^2=P=\int_0^\infty L(\nu)d\nu$.  Here $\nu_{0}$ is the Lorentzian centroid
    frequency and FWHM the full-width-at-half-maximum. A feature is called a
    QPO if $Q > 2$. Features with $Q<2$ are called noise
    components. We consider a component to be significant if the ratio of the
    integrated power to its single trial negative error is greater than
    three; $ P / {\sigma_{P}} \geq 3$. Boundary cases with
    $P\simeq3\sigma_P$ for which we deviate from this rule are explicitly
    discussed in Section \ref{sec:results}.
	
    In the 2008 and 2011 outbursts we previously found a broad noise
    component centered at 1--5~Hz \citep{Bult2014}, which we called \textit{high luminosity
    flaring}. This flaring component cannot be fitted with a Lorentzian
    profile, but instead is well described by a Schechter function, which is
    a power-law with an exponential cut-off, 
	$
		S(\nu)\propto \nu^{-\alpha} e^{-\nu/\nu_{\cut}}
	$
	\citep{Hasinger1989, Dotani1989}, with power law index $\alpha$
	and cut-off frequency $\nu_{\cut}$.
    To compare the Schechter function with Lorentzian components, and by analogy
    to the definitions used for Lorentzian components, we define the
    Schechter ``centroid'' frequency, $\nu_0 = -\alpha\nu_{\cut}$, as the frequency
    of maximum power in $S(\nu)$. Alternatively, we can
    define a characteristic frequency, $\nu_{\max} = (1-\alpha)\nu_{\cut}$, as
    the frequency of the maximum in $\nu S(\nu)$.

\section{Source State Identification}	
    Atoll sources are named after the shape they trace out in the
    color-color diagram \citep{Hasinger1989,Klis2006}. At high luminosities they exhibit a banana shaped
    pattern, which is divided into three regions corresponding to three 
    source states (\textit{upper, lower} and \textit{lower-left banana}). 
    At lower luminosities the hard color
    increases and the observations cluster in the \textit{island state}. In the
    hardest, lowest luminosity states a change in variability distinguishes
    another source state; the \textit{extreme island} state. Depending on the
    source state, atoll sources show a different number of power spectral
    features and have a different morphology (see \citealt{Klis2006} for a
    detailed overview).
	
    SAX~J1808 is a low luminosity atoll source \citep{Straaten2005} and has shown 
    only three source states: the extreme island state, the island 
    state and the lower-left banana. We discuss the power spectral shapes 
    relevant to these source states.

    In the \textit{extreme island state} three to four broad Lorentzians are needed to
    fit the power spectrum. The components with the lowest and second lowest characteristic 
    frequencies are respectively called the break ($b$) and hump ($h$) Lorentzians 
    \citep{Belloni2002}. The highest frequency component is usually found at 100--200~Hz and 
    is identified as ``$u$'' to represent the widely held idea that this component evolves into the
    upper kHz QPO found in the island state and the lower-left banana \citep{Psaltis1999}. 
    An additional component is usually present between the upper kHz and hump
    Lorentzian, which is proposed to evolve into the lower kHz QPO \citep{Psaltis1999}, but 
    that identification is debated \citep{Straaten2003}.
    In this work we follow \citet{Straaten2005} and label this feature as a separate
    ``$\low$'' component. The relation between the $\low$ component and the lower kHz
    QPO is further discussed in section~\ref{sec:ell-vs-low}

    In the \textit{island state} three or four Lorentzian profiles are needed. These components
    tend to be less broad, while covering a higher frequency range with respect
    to the Lorentzians seen in the extreme island state. The lowest frequency
    noise components are again called the break and hump Lorentzians. The
    highest frequency component is now a QPO, and identified as the upper
    kHz QPO \citep{Wijnands2003}. Often an additional Lorentzian is needed at $\nu_{\rm max}\sim100-200$~Hz, which is
    called the hectohertz (hHz) component \citep{Ford1998,Straaten2002}. This component
    is usually observed as a broad noise component with a quality factor below 0.5. Occasionally
    it becomes more peaked, but it always has $Q\lesssim2$. A characteristic property of the
    hHz component is that it does not show a systematic trend in frequency. It
    stays roughly constant in the hectohertz range. 
    A $\low$ component is not usually observed in the island state, but given their similar
    frequencies, the $\low$ and hHz components may be difficult to disentangle. We simply
    always label the noise component below the upper kHz QPO as the hHz component and discuss
    this issue further in section~\ref{sec.low-vs-hhz}.

    In the \textit{lower-left banana} the power spectra show a complex morphology and
    can require up to seven different Lorentzians. 
    In contrast to the two island states, the lower-left banana may show two breaks in the
    broad band noise, ``$b$'' and ``$b2$'', with ${b2}$ being the lower frequency
    component. As in the island state, higher frequency features are called the
    hump and hectohertz components. Finally, at the highest frequencies a lower 
    ($\ell$) and upper ($u$) kHz QPO is seen. If both kHz QPOs are detected in SAX~J1808, 
    the upper kHz QPO is always the most prominent feature, so if only one QPO is 
    present, it is identified as the upper kHz QPO \citep{Wijnands2003,Straaten2005}. 
    Because the lower kHz QPO always appear as a very sharp, weak QPO it can be easily 
    distinguished from the much broader hHz component that may be present at a lower frequency.
	
	In some atoll sources the hump feature occasionally shows two discernible peaks, or a
	complex morphology requiring two Lorentzian profiles for an acceptable fit. In such
	cases the narrow Lorentzian is called the Low-Frequency (LF) component, while the broader 
	Lorentzian is called the hump ($h$) component. In SAX~J1808 the two components are sometimes
	both narrow QPOs, in which case we call the lower frequency component LF and the higher
	frequency component the hump.
	
    Following \citet{Belloni2002} we refer to each Lorentzian profile as $L_i$,
    where subscript $i=\{b2,b,\LF,h,\low,\hHz,\ell,u\}$ represents the name of 
    the specific component (e.g. $L_b$ refers to the break component). We may 
    further use $\nu_{i}$ to refer to characteristic frequency and $\nu_{0,i}$ for 
    the centroid frequency. Similarly, the quality factor and fractional rms are 
    referred to as $Q_i$ and $r_i$.

\section{Results} \label{sec:results}
\begin{figure}[t]
		\includegraphics[width=0.9\linewidth]{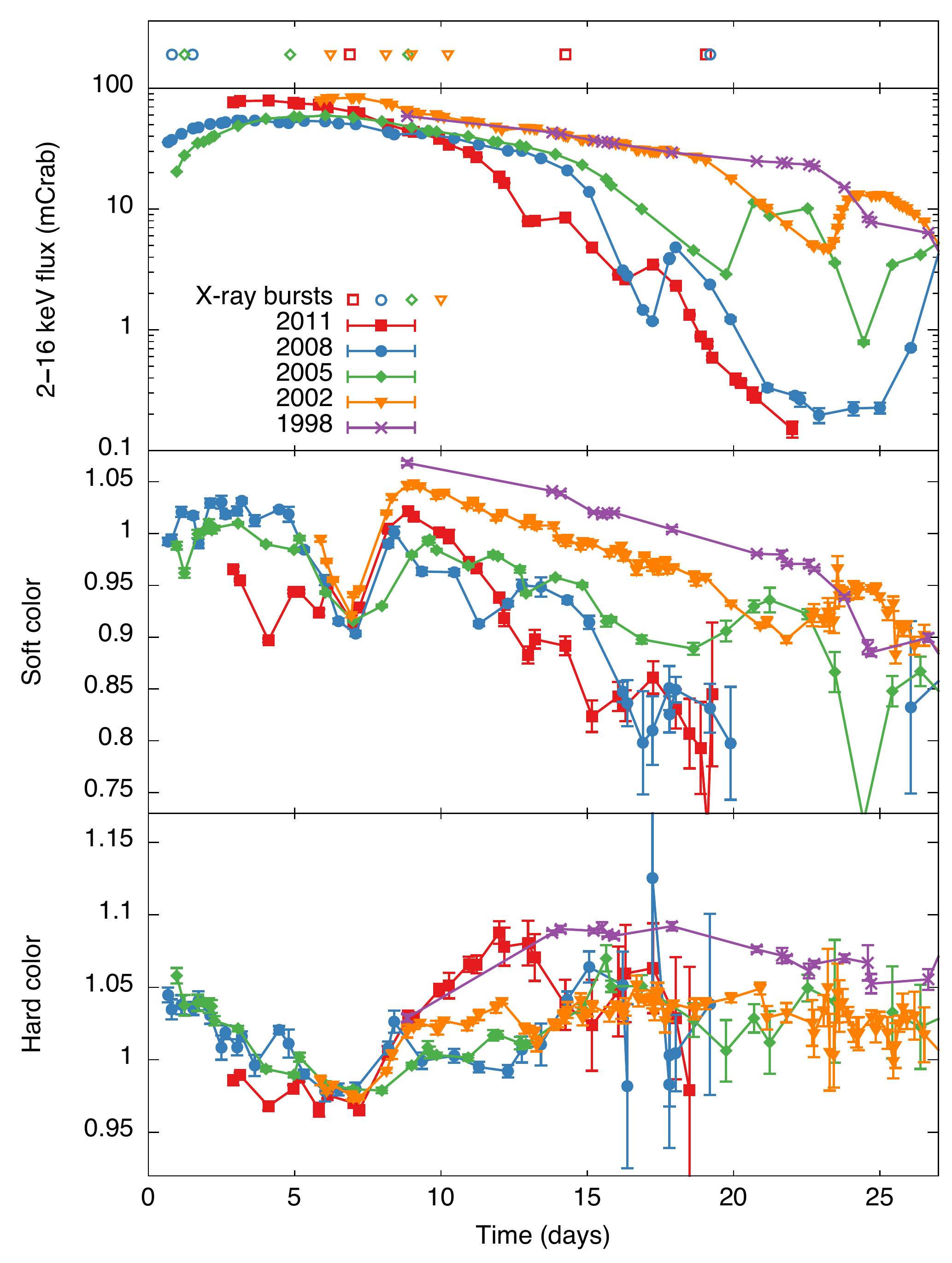}
        \caption{ 
			Evolution of the outbursts of SAX~J1808. The top panel gives the times 
			of type I X-ray bursts. The second	panel gives the 2--16 keV light curves, while 
			the third and fourth panel give the evolution of the soft and hard colors, 
			respectively. Color measurements with large uncertainties have been omitted for 
			clarity. Each light curve has been shifted in time by: 2011 MJD~55867.03; 2008 
			MJD~54731.20; 2005 MJD~53521.79; 2002 MJD~52556.19; and 1998 MJD~50906.03,
			such that their color minima approximately match on day 7.0. Each point in this figure 
			corresponds to an ObsID in Table~\ref{tab:intervals}.
		}
		\label{fig:OutburstEvolution}
	\end{figure}

\begin{figure*}[p]
		\centering
		\includegraphics[width=\linewidth]{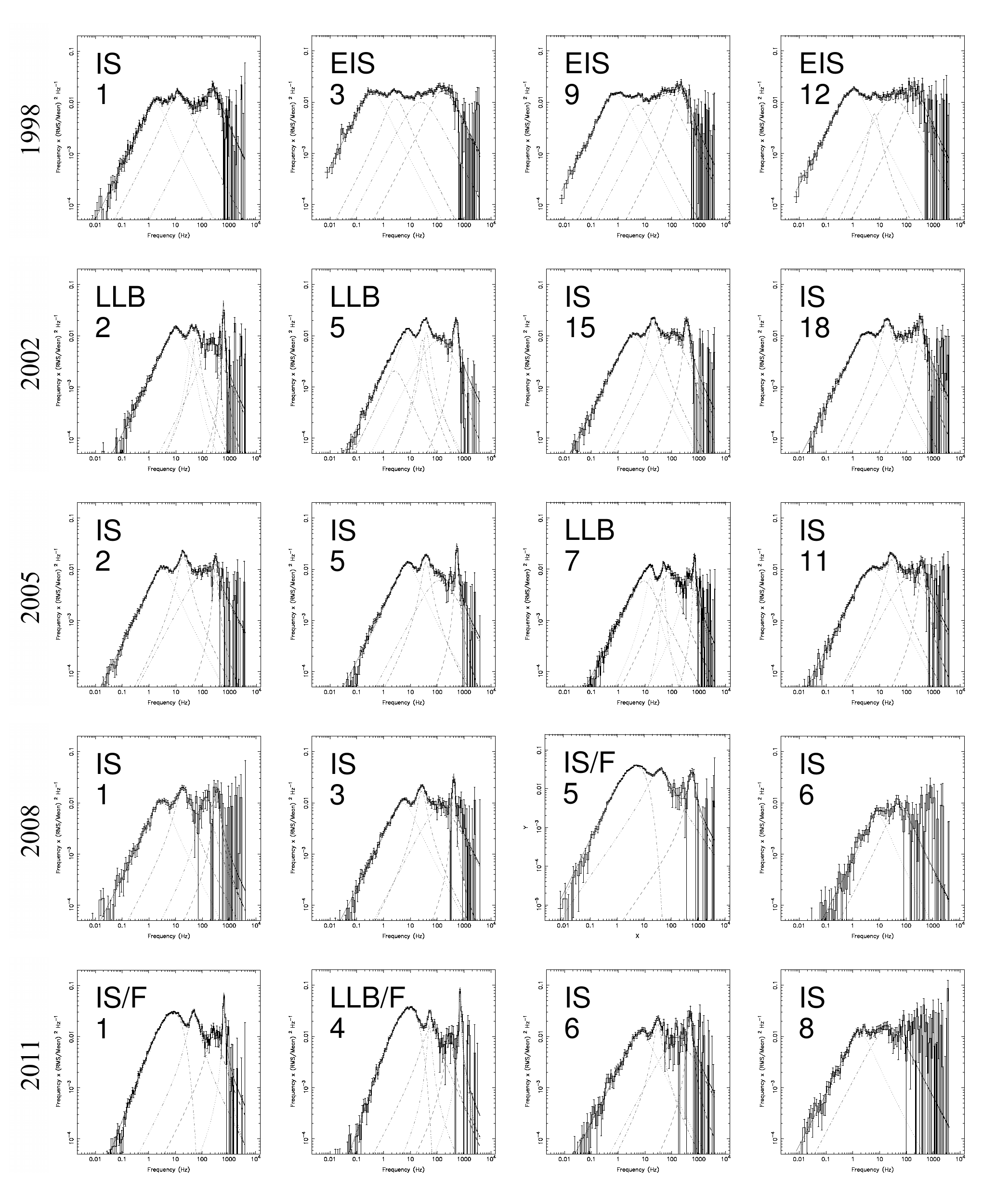}
		\caption{
			Selected power spectra of each outburst of SAX~J1808. Every
			row represents a separate outburst as indicated. Each power
			spectrum has been marked with the number of the interval in 
			its respective outburst and its associated source state
			(see Table~\ref{tab:intervals} for details). Parameters of 
			the fitted curves are given in 
			Tables~\ref{tab:Frequencies}--\ref{tab:Schechter}.
		}
		\label{fig:PowerSpectra}
	\end{figure*}

    We briefly discuss the timing and color evolution of SAX~J1808 for each
    individual outburst. We make use of Figure~\ref{fig:OutburstEvolution},
    which shows the 2--16 keV light curves, and the soft and hard color evolution
    for all outbursts. Additionally, we refer to Figure~\ref{fig:PowerSpectra},
    which shows a sample of representative power spectra for each outburst, and
    Tables~\ref{tab:Frequencies},~\ref{tab:Qualities}, \ref{tab:Amplitudes} and 
    \ref{tab:Schechter} which list the full set of fit parameters obtained for SAX~J1808. 
	
%
%
\begin{deluxetable*}{lcccccccc}
\tabletypesize{\scriptsize}
\tablecaption{
    Characteristic Frequency; $\nu_{\max}$~(Hz)
    \label{tab:Frequencies}
}
\tablewidth{ \linewidth }
\tablehead{
    \colhead{Interval} & \colhead{$L_{b2}$} & \colhead{$L_{b}$} & 
    \colhead{$L_h$} & \colhead{$L_{\LF}$} & \colhead{$L_{\low}$} & 
    \colhead{$L_{\hHz}$} & \colhead{$L_{\ell}$} & \colhead{$L_{u}$}
}
\startdata
\cutinhead{1998 Outburst}
1  &   \nodata   & $ 1.83  \pm  0.08  $ & $ 13.3 \pm  0.5  $ &   \nodata   &      \nodata      &   \nodata   &   \nodata   & $ 236 \pm  14 $ \\
2  &   \nodata   & $ 0.408 \pm  0.013 $ & $ 2.89 \pm  0.08 $ &   \nodata   & $ 26   \pm  2   $ &   \nodata   &   \nodata   & $ 155 \pm   7 $ \\
3  &   \nodata   & $ 0.381 \pm  0.018 $ & $ 2.79 \pm  0.10 $ &   \nodata   & $ 24   \pm  4   $ &   \nodata   &   \nodata   & $ 167 \pm  13 $ \\
4  &   \nodata   & $ 0.363 \pm  0.012 $ & $ 2.62 \pm  0.07 $ &   \nodata   & $ 21.9 \pm  1.8 $ &   \nodata   &   \nodata   & $ 150 \pm   6 $ \\
5  &   \nodata   & $ 0.38  \pm  0.03  $ & $ 2.39 \pm  0.16 $ &   \nodata   & $ 22   \pm  3   $ &   \nodata   &   \nodata   & $ 162 \pm  20 $ \\
6  &   \nodata   & $ 0.347 \pm  0.013 $ & $ 2.57 \pm  0.09 $ &   \nodata   & $ 28   \pm  3   $ &   \nodata   &   \nodata   & $ 177 \pm  11 $ \\
7  &   \nodata   & $ 0.359 \pm  0.014 $ & $ 2.63 \pm  0.09 $ &   \nodata   & $ 21   \pm  3   $ &   \nodata   &   \nodata   & $ 164 \pm  11 $ \\
8  &   \nodata   & $ 0.37  \pm  0.02  $ & $ 2.80 \pm  0.17 $ &   \nodata   & $ 27   \pm  6   $ &   \nodata   &   \nodata   & $ 159 \pm  16 $ \\
9  &   \nodata   & $ 0.81  \pm  0.04  $ & $ 5.5  \pm  0.3  $ &   \nodata   & $ 54   \pm  8   $ &   \nodata   &   \nodata   & $ 213 \pm  14 $ \\
10 &   \nodata   & $ 1.31  \pm  0.07  $ & $ 7.6  \pm  0.4  $ &   \nodata   & $ 52   \pm  8   $ &   \nodata   &   \nodata   & $ 199 \pm  19 $ \\
11 &   \nodata   & $ 2.01  \pm  0.08  $ & $ 13.1 \pm  0.8  $ &   \nodata   &      \nodata      &   \nodata   &   \nodata   & $ 170 \pm  14^{1} $ \\
12 &   \nodata   & $ 1.03  \pm  0.03  $ & $ 6.5  \pm  0.4  $ &   \nodata   & $ 32   \pm  10  $ &   \nodata   &   \nodata   & $ 193 \pm  34 $ \\
13 &   \nodata   & $ 0.93  \pm  0.03  $ &       \nodata      &   \nodata   & $ 37   \pm  4   $ &   \nodata   &   \nodata   &    \nodata    \\
\cutinhead{2002 Outburst}
1  &      \nodata      & $ 10.4  \pm  0.5  $ & $ 43.9  \pm  2.1  $ &      \nodata      &   \nodata   & $ 179 \pm  26 $ &     \nodata    & $ 567 \pm  25 $ \\
2  &      \nodata      & $  9.9  \pm  0.2  $ & $ 57.8  \pm  2.2  $ & $ 37.4 \pm  1.1 $ &   \nodata   & $ 212 \pm  34 $ & $ 435 \pm 12 $ & $ 599 \pm   5 $ \\
3  & $  7.9 \pm  1.5 $ & $ 15.4  \pm  0.4  $ & $ 78.2  \pm  2.0  $ & $ 47.2 \pm  1.8 $ &   \nodata   & $ 257 \pm  35 $ &     \nodata    & $ 698 \pm   6 $ \\
4  & $ 10.0 \pm  1.0 $ & $ 15.9  \pm  0.2  $ & $ 75.7  \pm  1.0  $ & $ 47.4 \pm  0.7 $ &   \nodata   & $ 355 \pm  43 $ & $ 504 \pm  3 $ & $ 686 \pm   5 $ \\
5  & $  2.6 \pm  0.5 $ & $  7.6  \pm  0.2  $ & $ 39.3  \pm  1.1  $ & $ 28.1 \pm  1.1 $ &   \nodata   & $ 101 \pm  13 $ &     \nodata    & $ 497 \pm   8 $ \\
6  &      \nodata      & $  3.14 \pm  0.05 $ & $ 18.01 \pm  0.15 $ &      \nodata      &   \nodata   & $ 122 \pm   9 $ &     \nodata    & $ 352 \pm   5 $ \\
7  &      \nodata      & $  3.08 \pm  0.05 $ & $ 18.53 \pm  0.19 $ &      \nodata      &   \nodata   & $ 122 \pm  11 $ &     \nodata    & $ 339 \pm   5 $ \\
8  &      \nodata      & $  3.00 \pm  0.05 $ & $ 18.40 \pm  0.19 $ &      \nodata      &   \nodata   & $ 127 \pm  12 $ &     \nodata    & $ 332 \pm   6 $ \\
9  &      \nodata      & $  2.26 \pm  0.05 $ & $ 13.6  \pm  0.3  $ &      \nodata      &   \nodata   & $ 112 \pm  12 $ &     \nodata    & $ 294 \pm   7 $ \\
10 &      \nodata      & $  4.6  \pm  0.2  $ & $ 24.4  \pm  1.0  $ &      \nodata      &   \nodata   & $  98 \pm  48 $ &     \nodata    & $ 389 \pm  21 $ \\
11 &      \nodata      & $  5.00 \pm  0.12 $ & $ 25.4  \pm  0.3  $ &      \nodata      &   \nodata   & $ 113 \pm  14 $ &     \nodata    & $ 393 \pm   5 $ \\
12 &      \nodata      & $  4.9  \pm  0.2  $ & $ 25.1  \pm  0.6  $ &      \nodata      &   \nodata   & $ 100 \pm  25 $ &     \nodata    & $ 381 \pm  10 $ \\
13 &      \nodata      & $  4.26 \pm  0.11 $ & $ 23.0  \pm  0.3  $ &      \nodata      &   \nodata   & $ 120 \pm  19 $ &     \nodata    & $ 370 \pm   7 $ \\
14 &      \nodata      & $  3.41 \pm  0.17 $ & $ 19.3  \pm  0.5  $ &      \nodata      &   \nodata   & $  89 \pm  28 $ &     \nodata    & $ 332 \pm  11 $ \\
15 &      \nodata      & $  3.71 \pm  0.10 $ & $ 20.2  \pm  0.3  $ &      \nodata      &   \nodata   & $ 109 \pm  10 $ &     \nodata    & $ 358 \pm  10 $ \\
16 &      \nodata      & $  3.53 \pm  0.19 $ & $ 23.2  \pm  0.7  $ &      \nodata      &   \nodata   & $  88 \pm   7 $ &     \nodata    & $ 336 \pm  17 $ \\
17 &      \nodata      & $  3.41 \pm  0.10 $ & $ 20.9  \pm  0.4  $ &      \nodata      &   \nodata   & $ 127 \pm  22 $ &     \nodata    & $ 326 \pm  10 $ \\
18 &      \nodata      & $  3.15 \pm  0.07 $ & $ 19.2  \pm  0.3  $ &      \nodata      &   \nodata   & $ 139 \pm  27 $ &     \nodata    & $ 340 \pm  11 $ \\
19 &      \nodata      & $  3.49 \pm  0.14 $ & $ 20.8  \pm  0.5  $ &      \nodata      &   \nodata   & $ 187 \pm  48 $ &     \nodata    & $ 358 \pm  14 $ \\
20 &      \nodata      & $  3.73 \pm  0.12 $ & $ 21.2  \pm  0.4  $ &      \nodata      &   \nodata   & $ 139 \pm  18 $ &     \nodata    & $ 332 \pm  11 $ \\
21 &      \nodata      & $  3.99 \pm  0.15 $ & $ 23.1  \pm  0.5  $ &      \nodata      &   \nodata   & $ 173 \pm  51 $ &     \nodata    & $ 357 \pm  16 $ \\
22 &      \nodata      & $  4.8  \pm  0.4  $ & $ 28.7  \pm  1.0  $ &      \nodata      &   \nodata   &     \nodata     &     \nodata    & $ 413 \pm 143 $ \\
\cutinhead{2005 Outburst}
1  &     \nodata     & $  2.6  \pm  0.2  $ & $ 23.7 \pm  1.9 $ &      \nodata      &   \nodata   &      \nodata      &   \nodata   &   \nodata   \\
2  &     \nodata     & $  3.15 \pm  0.06 $ & $ 18.9 \pm  0.2 $ &      \nodata      &   \nodata   & $ 122 \pm 21    $ &   \nodata   & $ 304 \pm 10 $ \\
3  &     \nodata     & $  5.51 \pm  0.12 $ & $ 28.1 \pm  0.4 $ &      \nodata      &   \nodata   & $  90 \pm 10    $ &   \nodata   & $ 399 \pm  7 $ \\
4  &     \nodata     & $  7.50 \pm  0.14 $ & $ 38.5 \pm  0.6 $ &      \nodata      &   \nodata   & $ 155 \pm 35    $ &   \nodata   & $ 526 \pm  8 $ \\
5  &     \nodata     & $  7.63 \pm  0.12 $ & $ 38.4 \pm  0.6 $ &      \nodata      &   \nodata   & $ 121 \pm 21    $ &   \nodata   & $ 532 \pm  5 $ \\
6  &     \nodata     & $  6.67 \pm  0.17 $ & $ 34.2 \pm  0.7 $ &      \nodata      &   \nodata   & $ 161 \pm 45    $ &   \nodata   & $ 494 \pm 14 $ \\
7  & $ 8.9 \pm 1.0 $ & $ 15.4  \pm  0.3  $ & $ 78   \pm  4   $ & $ 47.9 \pm  0.6 $ &   \nodata   & $ 438 \pm 56^{1}$ &   \nodata   & $ 685 \pm 10 $ \\
8  &     \nodata     & $  7.24 \pm  0.15 $ & $ 36.6 \pm  0.5 $ &      \nodata      &   \nodata   & $ 136 \pm 12    $ &   \nodata   & $ 498 \pm 11 $ \\
9  & $ 3.3 \pm 0.7 $ & $  7.9  \pm  0.3  $ & $ 36.5 \pm  0.4 $ &      \nodata      &   \nodata   & $ 131 \pm 19    $ &   \nodata   & $ 488 \pm  9 $ \\
10 &     \nodata     & $  6.70 \pm  0.16 $ & $ 34.4 \pm  0.4 $ &      \nodata      &   \nodata   & $ 127 \pm 27    $ &   \nodata   & $ 481 \pm 19 $ \\
11 &     \nodata     & $  5.07 \pm  0.18 $ & $ 28.3 \pm  0.6 $ &      \nodata      &   \nodata   & $ 138 \pm 30    $ &   \nodata   & $ 393 \pm 25 $ \\
\cutinhead{2008 Outburst}
1  &   \nodata   & $ 2.85 \pm  0.10 $ & $ 18.7 \pm  0.8 $ &   \nodata   &   \nodata   & $ 135 \pm  17 $ &   \nodata   & $ 350 \pm  36 $ \\
2  &   \nodata   & $ 3.66 \pm  0.17 $ & $ 19.5 \pm  0.4 $ &   \nodata   &   \nodata   & $ 118 \pm  23 $ &   \nodata   & $ 372 \pm  14 $ \\
3  &   \nodata   & $ 5.35 \pm  0.15 $ & $ 26.5 \pm  0.5 $ &   \nodata   &   \nodata   & $ 135 \pm  29 $ &   \nodata   & $ 415 \pm   5 $ \\
4  &   \nodata   & $ 9.2  \pm  0.5  $ & $ 49   \pm  2   $ &   \nodata   &   \nodata   &     \nodata     &   \nodata   & $ 524 \pm  33 $ \\
5  &   \nodata   &      --- $^{2}$    & $ 37   \pm  3   $ &   \nodata   &   \nodata   &     \nodata     &   \nodata   & $ 606 \pm  44 $ \\
6  &   \nodata   & $ 8.7  \pm  1.3  $ & $ 54   \pm 10   $ &   \nodata   &   \nodata   &     \nodata     &   \nodata   &   \nodata   \\
7  &   \nodata   & $ 1.72 \pm  0.16 $ & $ 23   \pm  4   $ &   \nodata   &   \nodata   &     \nodata     &   \nodata   &   \nodata   \\
\cutinhead{2011 Outburst}
1  &      \nodata     &     --- $^{2}$      & $ 49.1 \pm  1.4 $ &      \nodata      &   \nodata   & $ 291 \pm 203 $ &    \nodata    & $ 657 \pm   3 $ \\
2  &      \nodata     &     --- $^{2}$      & $ 38   \pm 16   $ &      \nodata      &   \nodata   &     \nodata     &    \nodata    & $ 527 \pm 226 $ \\
3  &      \nodata     &     --- $^{2}$      & $ 52   \pm  8   $ & $ 52.3 \pm  0.7 $ &   \nodata   & $ 343 \pm  22 $ &    \nodata    & $ 719 \pm   4 $ \\
4  &      \nodata     &     --- $^{2}$      & $ 59   \pm 14   $ & $ 51.6 \pm  1.5 $ &   \nodata   & $ 289 \pm  43 $ &    \nodata    & $ 712 \pm   5 $ \\
5  & $ 4.4 \pm  1.1 $ & $ 15.3  \pm  0.8  $ & $ 78   \pm  4   $ & $ 48.8 \pm  1.7 $ &   \nodata   &     \nodata     & $ 567 \pm 9 $ & $ 737 \pm  10 $ \\
6  &      \nodata     & $  6.8  \pm  0.4  $ & $ 30.3 \pm  1.0 $ &      \nodata      &   \nodata   & $ 153 \pm  71 $ &    \nodata    & $ 478 \pm  25 $ \\
7  &      \nodata     & $  3.7  \pm  0.3  $ & $ 21.1 \pm  0.9 $ &      \nodata      &   \nodata   &     \nodata     &    \nodata    & $ 245 \pm  41 $ \\
8  &      \nodata     & $  1.90 \pm  0.13 $ & $ 22   \pm  3   $ &      \nodata      &   \nodata   &     \nodata     &    \nodata    &   \nodata   \\
9  &      \nodata     & $  0.55 \pm  0.04 $ & $ 4.6  \pm  1.2 $ &      \nodata      &   \nodata   &     \nodata     &    \nodata    & $ 113 \pm  36 $ \\
10 &      \nodata     & $  0.50 \pm  0.04 $ & $ 10   \pm  7   $ &      \nodata      &   \nodata   &     \nodata     &    \nodata    &   \nodata   \\
\enddata
\tablenotetext{1}{Uncertain identification.}
\tablenotetext{2}{Schechter function fit to the high luminosity flaring, see Table~\ref{tab:Schechter}.}
\end{deluxetable*}

%
%
\begin{deluxetable*}{lcccccccc}
\tabletypesize{\scriptsize}
\tablecaption{
	Quality Factor; $Q$
    \label{tab:Qualities}
}
\tablewidth{0pt}
\tablehead{
    \colhead{Interval} & \colhead{$L_{b2}$} & \colhead{$L_{b}$} & 
    \colhead{$L_h$} & \colhead{$L_{\LF}$} & \colhead{$L_{\low}$} & 
    \colhead{$L_{\hHz}$} & \colhead{$L_{\ell}$} & \colhead{$L_{u}$}
}
\startdata
\cutinhead{1998 Outburst}
1  &   \nodata   & $ 0.37 \pm  0.04 $ & $ 0.29 \pm 0.07 $ &   \nodata   &      \nodata      &   \nodata   &   \nodata   & $ 0.43 \pm  0.10 $ \\
2  &   \nodata   & $ 0.21 \pm  0.03 $ & $ 0.34 \pm 0.06 $ &   \nodata   & $ 0.11 \pm 0.14 $ &   \nodata   &   \nodata   & $ 0.35 \pm  0.08 $ \\
3  &   \nodata   & $ 0.19 \pm  0.04 $ & $ 0.41 \pm 0.08 $ &   \nodata   &    0 ( fixed )    &   \nodata   &   \nodata   & $ 0.24 \pm  0.10 $ \\
4  &   \nodata   & $ 0.18 \pm  0.03 $ & $ 0.41 \pm 0.06 $ &   \nodata   &    0 ( fixed )    &   \nodata   &   \nodata   & $ 0.38 \pm  0.07 $ \\
5  &   \nodata   & $ 0.25 \pm  0.06 $ & $ 0.43 \pm 0.12 $ &   \nodata   &    0 ( fixed )    &   \nodata   &   \nodata   & $ 0.32 \pm  0.16 $ \\
6  &   \nodata   & $ 0.22 \pm  0.03 $ & $ 0.23 \pm 0.06 $ &   \nodata   &    0 ( fixed )    &   \nodata   &   \nodata   & $ 0.30 \pm  0.10 $ \\
7  &   \nodata   & $ 0.23 \pm  0.03 $ & $ 0.35 \pm 0.08 $ &   \nodata   &    0 ( fixed )    &   \nodata   &   \nodata   & $ 0.21 \pm  0.11 $ \\
8  &   \nodata   & $ 0.28 \pm  0.05 $ & $ 0.24 \pm 0.12 $ &   \nodata   &    0 ( fixed )    &   \nodata   &   \nodata   & $ 0.41 \pm  0.19 $ \\
9  &   \nodata   & $ 0.20 \pm  0.03 $ & $ 0.28 \pm 0.13 $ &   \nodata   & $ 0.19 \pm 0.16 $ &   \nodata   &   \nodata   & $ 0.8  \pm  0.2  $ \\
10 &   \nodata   & $ 0.16 \pm  0.03 $ & $ 0.5  \pm 0.2  $ &   \nodata   & $ 0.4  \pm 0.2  $ &   \nodata   &   \nodata   & $ 0.5  \pm  0.2  $ \\
11 &   \nodata   & $ 0.12 \pm  0.02 $ & $ 0.57 \pm 0.19 $ &   \nodata   &      \nodata      &   \nodata   &   \nodata   &  0 ( fixed )$^{1}$ \\
12 &   \nodata   & $ 0.15 \pm  0.02 $ & $ 1.1  \pm 0.5  $ &   \nodata   &    0 ( fixed )    &   \nodata   &   \nodata   & $ 0.4  \pm  0.3  $ \\
13 &   \nodata   & $ 0.02 \pm  0.03 $ &      \nodata      &   \nodata   &    0 ( fixed )    &   \nodata   &   \nodata   &   \nodata   \\
\cutinhead{2002 Outburst}
1  &       \nodata      & $ 0.35  \pm  0.04  $ & $ 0.89 \pm 0.18 $ &     \nodata     &   \nodata   & $ 1.2  \pm 0.7  $ &     \nodata    &   0 ( fixed )   \\
2  &       \nodata      & $ 0.35  \pm  0.02  $ & $ 1.9  \pm 0.7  $ & $ 2.4 \pm 0.6 $ &   \nodata   & $ 0.4  \pm 0.3  $ & $ 12 \pm 113 $ & $ 7.7 \pm 1.7 $ \\
3  & $ 0.35 \pm  0.06 $ & $ 1.3   \pm  0.3   $ & $ 2.8  \pm 0.8  $ & $ 2.0 \pm 0.5 $ &   \nodata   & $ 0.6  \pm 0.3  $ &     \nodata    & $ 8.2 \pm 1.9 $ \\
4  & $ 0.28 \pm  0.02 $ & $ 1.3   \pm  0.2   $ & $ 3.4  \pm 0.6  $ & $ 2.3 \pm 0.3 $ &   \nodata   &    0 ( fixed )    &  27 ( fixed )  & $ 7.0 \pm 1.2 $ \\
5  & $ 0.36 \pm  0.07 $ & $ 0.58  \pm  0.10  $ & $ 1.9  \pm 0.4  $ & $ 1.9 \pm 0.3 $ &   \nodata   &    0 ( fixed )    &     \nodata    & $ 3.0 \pm 0.3 $ \\
6  &       \nodata      & $ 0.236 \pm  0.012 $ & $ 0.85 \pm 0.04 $ &     \nodata     &   \nodata   & $ 0.19 \pm 0.08 $ &     \nodata    & $ 2.3 \pm 0.3 $ \\
7  &       \nodata      & $ 0.260 \pm  0.014 $ & $ 0.83 \pm 0.04 $ &     \nodata     &   \nodata   & $ 0.20 \pm 0.11 $ &     \nodata    & $ 2.3 \pm 0.4 $ \\
8  &       \nodata      & $ 0.279 \pm  0.013 $ & $ 0.86 \pm 0.05 $ &     \nodata     &   \nodata   & $ 0.24 \pm 0.12 $ &     \nodata    & $ 2.4 \pm 0.5 $ \\
9  &       \nodata      & $ 0.217 \pm  0.016 $ & $ 0.65 \pm 0.06 $ &     \nodata     &   \nodata   & $ 0.14 \pm 0.11 $ &     \nodata    & $ 1.9 \pm 0.5 $ \\
10 &       \nodata      & $ 0.33  \pm  0.04  $ & $ 1.1  \pm 0.3  $ &     \nodata     &   \nodata   &    0 ( fixed )    &     \nodata    & $ 2.8 \pm 1.8 $ \\
11 &       \nodata      & $ 0.267 \pm  0.018 $ & $ 1.18 \pm 0.10 $ &     \nodata     &   \nodata   & $ 0.15 \pm 0.19 $ &     \nodata    & $ 4.4 \pm 0.6 $ \\
12 &       \nodata      & $ 0.30  \pm  0.03  $ & $ 1.14 \pm 0.17 $ &     \nodata     &   \nodata   &    0 ( fixed )    &     \nodata    & $ 5   \pm 3   $ \\
13 &       \nodata      & $ 0.265 \pm  0.019 $ & $ 0.93 \pm 0.07 $ &     \nodata     &   \nodata   & $ 0.3  \pm 0.2  $ &     \nodata    & $ 2.8 \pm 0.8 $ \\
14 &       \nodata      & $ 0.27  \pm  0.04  $ & $ 0.98 \pm 0.16 $ &     \nodata     &   \nodata   &    0 ( fixed )    &     \nodata    & $ 4   \pm 2   $ \\
15 &       \nodata      & $ 0.240 \pm  0.018 $ & $ 0.99 \pm 0.08 $ &     \nodata     &   \nodata   & $ 0.32 \pm 0.16 $ &     \nodata    & $ 2.5 \pm 0.5 $ \\
16 &       \nodata      & $ 0.26  \pm  0.04  $ & $ 0.72 \pm 0.10 $ &     \nodata     &   \nodata   & $ 1.4  \pm 0.6  $ &     \nodata    & $ 1.5 \pm 0.4 $ \\
17 &       \nodata      & $ 0.255 \pm  0.021 $ & $ 0.88 \pm 0.09 $ &     \nodata     &   \nodata   & $ 0.3  \pm 0.2  $ &     \nodata    & $ 2.6 \pm 0.9 $ \\
18 &       \nodata      & $ 0.245 \pm  0.017 $ & $ 0.91 \pm 0.07 $ &     \nodata     &   \nodata   & $ 0.18 \pm 0.16 $ &     \nodata    & $ 2.0 \pm 0.7 $ \\
19 &       \nodata      & $ 0.25  \pm  0.03  $ & $ 0.85 \pm 0.10 $ &     \nodata     &   \nodata   &    0 ( fixed )    &     \nodata    & $ 3.7 \pm 1.9 $ \\
20 &       \nodata      & $ 0.24  \pm  0.02  $ & $ 1.01 \pm 0.09 $ &     \nodata     &   \nodata   & $ 0.34 \pm 0.19 $ &     \nodata    & $ 3.0 \pm 1.3 $ \\
21 &       \nodata      & $ 0.24  \pm  0.03  $ & $ 1.04 \pm 0.11 $ &     \nodata     &   \nodata   &    0 ( fixed )    &     \nodata    & $ 3.2 \pm 1.6 $ \\
22 &       \nodata      & $ 0.21  \pm  0.06  $ & $ 1.4  \pm 0.2  $ &     \nodata     &   \nodata   &      \nodata      &     \nodata    &   0 ( fixed )   \\
\cutinhead{2005 Outburst}
1  &      \nodata      & $ 0.49  \pm  0.12  $ & $ 0.7  \pm  0.2  $ &     \nodata     &   \nodata   &       \nodata        &   \nodata   &     \nodata      \\
2  &      \nodata      & $ 0.315 \pm  0.017 $ & $ 1.08 \pm  0.07 $ &     \nodata     &   \nodata   &    0 ( fixed )       &   \nodata   & $ 3.4 \pm  1.7 $ \\
3  &      \nodata      & $ 0.324 \pm  0.019 $ & $ 1.28 \pm  0.10 $ &     \nodata     &   \nodata   & $ 0.40 \pm  0.16 $   &   \nodata   & $ 4.5 \pm  0.9 $ \\
4  &      \nodata      & $ 0.308 \pm  0.016 $ & $ 1.31 \pm  0.11 $ &     \nodata     &   \nodata   &   0 ( fixed )        &   \nodata   & $ 4.4 \pm  1.5 $ \\
5  &      \nodata      & $ 0.319 \pm  0.014 $ & $ 1.37 \pm  0.12 $ &     \nodata     &   \nodata   &   0 ( fixed )        &   \nodata   & $ 5.8 \pm  1.9 $ \\
6  &      \nodata      & $ 0.41  \pm  0.03  $ & $ 1.6  \pm  0.2  $ &     \nodata     &   \nodata   &   0 ( fixed )        &   \nodata   & $ 9   \pm  6   $ \\
7  & $ 0.26 \pm 0.02 $ & $ 1.3   \pm  0.2   $ & $ 0.73 \pm  0.15 $ & $ 4.0 \pm 1.0 $ &   \nodata   & $ 0.8  \pm  0.5^{1}$ &   \nodata   & $ 7   \pm  6   $ \\
8  &      \nodata      & $ 0.293 \pm  0.018 $ & $ 1.63 \pm  0.12 $ &     \nodata     &   \nodata   & $ 0.53 \pm  0.18 $   &   \nodata   & $ 3.8 \pm  1.0 $ \\
9  & $ 0.34 \pm 0.04 $ & $ 0.69  \pm  0.11  $ & $ 1.30 \pm  0.08 $ &     \nodata     &   \nodata   & $ 0.24 \pm  0.15 $   &   \nodata   & $ 2.7 \pm  0.5 $ \\
10 &      \nodata      & $ 0.253 \pm  0.019 $ & $ 1.56 \pm  0.12 $ &     \nodata     &   \nodata   & $ 0.3  \pm  0.3  $   &   \nodata   & $ 2.3 \pm  0.9 $ \\
11 &      \nodata      & $ 0.18  \pm  0.02  $ & $ 0.95 \pm  0.11 $ &     \nodata     &   \nodata   &   0 ( fixed )        &   \nodata   & $ 2.1 \pm  0.8 $ \\
\cutinhead{2008 Outburst}
1  &   \nodata   & $ 0.35 \pm 0.04 $ & $ 0.79 \pm  0.09 $ &   \nodata   &   \nodata   & $ 1.0 \pm 0.4 $ &   \nodata   & $ 2.4 \pm 1.0 $ \\
2  &   \nodata   & $ 0.20 \pm 0.04 $ & $ 1.45 \pm  0.18 $ &   \nodata   &   \nodata   &   0 ( fixed )   &   \nodata   & $ 4   \pm 4   $ \\
3  &   \nodata   & $ 0.28 \pm 0.02 $ & $ 1.20 \pm  0.11 $ &   \nodata   &   \nodata   &   0 ( fixed )   &   \nodata   & $ 6   \pm 4   $ \\
4  &   \nodata   & $ 0.29 \pm 0.03 $ & $ 0.67 \pm  0.13 $ &   \nodata   &   \nodata   &     \nodata     &   \nodata   & $ 1.5 \pm 0.5 $ \\
5  &   \nodata   &      ---$^2$      & $ 0.46 \pm  0.05 $ &   \nodata   &   \nodata   &     \nodata     &   \nodata   & $ 2.2 \pm 0.7 $ \\
6  &   \nodata   & $ 0.35 \pm 0.17 $ & $ 0.3  \pm  0.5  $ &   \nodata   &   \nodata   &     \nodata     &   \nodata   &     \nodata     \\
7  &   \nodata   &    0 ( fixed )    &     0 ( fixed )    &   \nodata   &   \nodata   &     \nodata     &   \nodata   &     \nodata     \\
\cutinhead{2011 Outburst}
1  &      \nodata      &      ---$^2$       & $ 1.09 \pm  0.08 $ &      \nodata     &   \nodata   & $ 0.5 \pm 0.5 $ &   \nodata    & $  9   \pm 3   $ \\
2  &      \nodata      &      ---$^2$       & $ 0.4  \pm  0.3  $ &      \nodata     &   \nodata   &     \nodata     &   \nodata    & $  0.9 \pm 0.9 $ \\
3  &      \nodata      &      ---$^2$       & $ 0.42 \pm  0.10 $ & $ 5   \pm  3   $ &   \nodata   & $ 3.3 \pm 1.3 $ &   \nodata    & $  7.6 \pm 1.0 $ \\
4  &      \nodata      &      ---$^2$       & $ 0.60 \pm  0.18 $ & $ 3.3 \pm  1.3 $ &   \nodata   & $ 1.6 \pm 0.8 $ &   \nodata    & $  7.8 \pm 1.2 $ \\
5  & $ 0.50 \pm 0.13 $ & $ 0.54 \pm  0.12 $ & $ 1.9  \pm  0.9  $ & $ 4   \pm  4   $ &   \nodata   &     \nodata     & 14 ( fixed ) & $ 10   \pm 5   $ \\
6  &      \nodata      & $ 0.23 \pm  0.04 $ & $ 1.5  \pm  0.3  $ &      \nodata     &   \nodata   &   0 ( fixed )   &   \nodata    & $  3   \pm 2   $ \\
7  &      \nodata      & $ 0.27 \pm  0.05 $ & $ 1.0  \pm  0.2  $ &      \nodata     &   \nodata   &     \nodata     &   \nodata    & $  0.4 \pm 0.2 $ \\
8  &      \nodata      & $ 0.27 \pm  0.05 $ &     0 ( fixed )    &      \nodata     &   \nodata   &     \nodata     &   \nodata    &      \nodata     \\
9  &      \nodata      & $ 0.20 \pm  0.05 $ &     0 ( fixed )    &      \nodata     &   \nodata   &     \nodata     &   \nodata    &    0 ( fixed )   \\
10 &      \nodata      &     0 ( fixed )    &     0 ( fixed )    &      \nodata     &   \nodata   &     \nodata     &   \nodata    &      \nodata     \\
\end{deluxetable*}

%
%
\begin{deluxetable*}{lcccccccc}
\tabletypesize{\scriptsize}
\tablecaption{
	RMS Amplitudes; $r$~(\%)
	\label{tab:Amplitudes}
}
\tablewidth{0pt}
\tablehead{
    \colhead{Interval} & \colhead{$L_{b2}$} & \colhead{$L_{b}$} & 
    \colhead{$L_h$} & \colhead{$L_{\LF}$} & \colhead{$L_{\low}$} & 
    \colhead{$L_{\hHz}$} & \colhead{$L_{\ell}$} & \colhead{$L_{u}$}
}
\startdata
\cutinhead{1998 Outburst}
1  &   \nodata   & $ 14.4 \pm 0.5 $ & $ 17.8 \pm 0.7 $ &   \nodata   &     \nodata      &   \nodata   &   \nodata   & $ 18.6 \pm  0.7 $ \\
2  &   \nodata   & $ 19.0 \pm 0.3 $ & $ 16.6 \pm 0.8 $ &   \nodata   & $ 16.8 \pm 1.6 $ &   \nodata   &   \nodata   & $ 20.1 \pm  1.1 $ \\
3  &   \nodata   & $ 19.9 \pm 0.4 $ & $ 15.7 \pm 0.9 $ &   \nodata   & $ 17.7 \pm 0.9 $ &   \nodata   &   \nodata   & $ 20.9 \pm  1.2 $ \\
4  &   \nodata   & $ 20.5 \pm 0.3 $ & $ 15.5 \pm 0.6 $ &   \nodata   & $ 18.9 \pm 0.4 $ &   \nodata   &   \nodata   & $ 20.0 \pm  0.7 $ \\
5  &   \nodata   & $ 20.2 \pm 0.7 $ & $ 15.4 \pm 1.3 $ &   \nodata   & $ 19.0 \pm 1.0 $ &   \nodata   &   \nodata   & $ 19.2 \pm  1.5 $ \\
6  &   \nodata   & $ 19.6 \pm 0.4 $ & $ 17.5 \pm 0.7 $ &   \nodata   & $ 19.1 \pm 0.6 $ &   \nodata   &   \nodata   & $ 19.1 \pm  1.0 $ \\
7  &   \nodata   & $ 20.5 \pm 0.4 $ & $ 15.7 \pm 0.9 $ &   \nodata   & $ 18.3 \pm 0.7 $ &   \nodata   &   \nodata   & $ 21.0 \pm  1.1 $ \\
8  &   \nodata   & $ 19.3 \pm 0.7 $ & $ 17.0 \pm 1.3 $ &   \nodata   & $ 19.1 \pm 1.2 $ &   \nodata   &   \nodata   & $ 19.0 \pm  1.9 $ \\
9  &   \nodata   & $ 19.1 \pm 0.5 $ & $ 14.6 \pm 1.5 $ &   \nodata   & $ 19   \pm 2   $ &   \nodata   &   \nodata   & $ 16   \pm  2   $ \\
10 &   \nodata   & $ 18.6 \pm 0.6 $ & $ 11.8 \pm 1.7 $ &   \nodata   & $ 15   \pm 3   $ &   \nodata   &   \nodata   & $ 18   \pm  2   $ \\
11 &   \nodata   & $ 18.6 \pm 0.4 $ & $ 10.7 \pm 1.1 $ &   \nodata   &     \nodata      &   \nodata   &   \nodata   & $ 24.2 \pm  0.5^{1} $ \\
12 &   \nodata   & $ 21.5 \pm 0.2 $ & $  8.2 \pm 1.6 $ &   \nodata   & $ 19.0 \pm 1.8 $ &   \nodata   &   \nodata   & $ 18   \pm  3   $ \\
13 &   \nodata   & $ 25.4 \pm 0.3 $ &      \nodata     &   \nodata   & $ 28.3 \pm 0.8 $ &   \nodata   &   \nodata   &      \nodata   \\
\cutinhead{2002 Outburst}
1  &     \nodata     & $ 17.7  \pm  0.5  $ & $ 13.7 \pm 1.0 $ &     \nodata     &   \nodata   & $  8.4 \pm 1.7 $ &     \nodata     & $  8.0 \pm  1.2 $ \\
2  &     \nodata     & $ 17.4  \pm  0.2  $ & $  8.2 \pm 1.6 $ & $ 7.3 \pm 1.0 $ &   \nodata   & $ 11.8 \pm 1.6 $ & $ 4.0 \pm 1.0 $ & $  8.7 \pm  0.6 $ \\
3  & $ 9.6 \pm 1.3 $ & $  9.8  \pm  1.3  $ & $  7.1 \pm 1.1 $ & $ 7.3 \pm 0.9 $ &   \nodata   & $  9.9 \pm 1.1 $ &     \nodata     & $  8.3 \pm  0.7 $ \\
4  & $12.2 \pm 1.0 $ & $  9.0  \pm  1.1  $ & $  6.0 \pm 0.5 $ & $ 7.2 \pm 0.4 $ &   \nodata   & $ 12.8 \pm 0.4 $ & $ 2.5 \pm 0.4 $ & $  7.6 \pm  0.6 $ \\
5  & $ 6.8 \pm 1.7 $ & $ 14.2  \pm  1.0  $ & $  9.3 \pm 1.4 $ & $ 7.8 \pm 1.3 $ &   \nodata   & $ 15.5 \pm 0.5 $ &     \nodata     & $  9.7 \pm  0.4 $ \\
6  &     \nodata     & $ 15.36 \pm  0.12 $ & $ 15.3 \pm 0.3 $ &      \nodata    &   \nodata   & $ 15.7 \pm 0.7 $ &     \nodata     & $ 10.1 \pm  0.7 $ \\
7  &     \nodata     & $ 15.05 \pm  0.13 $ & $ 15.3 \pm 0.4 $ &      \nodata    &   \nodata   & $ 15.1 \pm 0.9 $ &     \nodata     & $ 10.6 \pm  0.8 $ \\
8  &     \nodata     & $ 15.17 \pm  0.12 $ & $ 15.6 \pm 0.4 $ &      \nodata    &   \nodata   & $ 15.0 \pm 0.9 $ &     \nodata     & $ 10.0 \pm  0.9 $ \\
9  &     \nodata     & $ 16.08 \pm  0.18 $ & $ 14.0 \pm 0.6 $ &      \nodata    &   \nodata   & $ 17.4 \pm 1.1 $ &     \nodata     & $ 10.4 \pm  1.3 $ \\
10 &     \nodata     & $ 15.24 \pm  0.4  $ & $ 14.1 \pm 1.7 $ &      \nodata    &   \nodata   & $ 15.6 \pm 1.3 $ &     \nodata     & $ 10   \pm  2   $ \\
11 &     \nodata     & $ 15.7  \pm  0.2  $ & $ 14.0 \pm 0.7 $ &      \nodata    &   \nodata   & $ 14.7 \pm 1.2 $ &     \nodata     & $  9.8 \pm  0.6 $ \\
12 &     \nodata     & $ 15.4  \pm  0.3  $ & $ 13.7 \pm 1.0 $ &      \nodata    &   \nodata   & $ 15.9 \pm 0.8 $ &     \nodata     & $  8.9 \pm  1.1 $ \\
13 &     \nodata     & $ 15.2  \pm  0.2  $ & $ 15.9 \pm 0.6 $ &      \nodata    &   \nodata   & $ 14.0 \pm 1.7 $ &     \nodata     & $ 10.8 \pm  1.2 $ \\
14 &     \nodata     & $ 14.7  \pm  0.3  $ & $ 14.5 \pm 1.1 $ &      \nodata    &   \nodata   & $ 15.7 \pm 0.9 $ &     \nodata     & $  9.2 \pm  1.4 $ \\
15 &     \nodata     & $ 15.44 \pm  0.19 $ & $ 15.1 \pm 0.6 $ &      \nodata    &   \nodata   & $ 14.8 \pm 1.1 $ &     \nodata     & $ 10.3 \pm  0.9 $ \\
16 &     \nodata     & $ 15.0  \pm  0.4  $ & $ 17.5 \pm 0.7 $ &      \nodata    &   \nodata   & $  9.1 \pm 1.5 $ &     \nodata     & $ 14.2 \pm  1.2 $ \\
17 &     \nodata     & $ 15.4  \pm  0.2  $ & $ 15.7 \pm 0.7 $ &      \nodata    &   \nodata   & $ 14.5 \pm 1.5 $ &     \nodata     & $  9.8 \pm  1.4 $ \\
18 &     \nodata     & $ 16.15 \pm  0.18 $ & $ 15.3 \pm 0.5 $ &      \nodata    &   \nodata   & $ 15.8 \pm 1.6 $ &     \nodata     & $ 11.0 \pm  1.8 $ \\
19 &     \nodata     & $ 15.7  \pm  0.3  $ & $ 15.3 \pm 0.6 $ &      \nodata    &   \nodata   & $ 17.1 \pm 1.3 $ &     \nodata     & $  8.9 \pm  1.6 $ \\
20 &     \nodata     & $ 15.6  \pm  0.2  $ & $ 15.6 \pm 0.6 $ &      \nodata    &   \nodata   & $ 15.2 \pm 1.4 $ &     \nodata     & $  9.2 \pm  1.5 $ \\
21 &     \nodata     & $ 15.9  \pm  0.3  $ & $ 15.0 \pm 0.7 $ &      \nodata    &   \nodata   & $ 17.2 \pm 1.3 $ &     \nodata     & $  8.8 \pm  1.9 $ \\
22 &     \nodata     & $ 15.7  \pm  0.5  $ & $ 14.3 \pm 0.9 $ &      \nodata    &   \nodata   &     \nodata      &     \nodata     & $ 21   \pm  2   $ \\
\cutinhead{2005 Outburst}
1  &     \nodata     & $ 13.5  \pm 0.8  $ & $ 18.4 \pm 1.3 $ &     \nodata     &   \nodata   &      \nodata        &   \nodata   &      \nodata     \\
2  &     \nodata     & $ 15.28 \pm 0.14 $ & $ 14.5 \pm 0.4 $ &     \nodata     &   \nodata   & $ 16.2 \pm 0.7 $    &   \nodata   & $  7.4 \pm 1.3 $ \\
3  &     \nodata     & $ 14.72 \pm 0.18 $ & $ 13.0 \pm 0.6 $ &     \nodata     &   \nodata   & $ 12.4 \pm 1.0 $    &   \nodata   & $  8.7 \pm 0.7 $ \\
4  &     \nodata     & $ 16.47 \pm 0.16 $ & $ 11.9 \pm 0.5 $ &     \nodata     &   \nodata   & $ 13.1 \pm 0.6 $    &   \nodata   & $  8.5 \pm 0.8 $ \\
5  &     \nodata     & $ 16.72 \pm 0.15 $ & $ 11.6 \pm 0.6 $ &     \nodata     &   \nodata   & $ 13.8 \pm 0.6 $    &   \nodata   & $  8.8 \pm 0.5 $ \\
6  &     \nodata     & $ 16.6  \pm 0.2  $ & $ 12.1 \pm 0.7 $ &     \nodata     &   \nodata   & $ 14.3 \pm 0.9 $    &   \nodata   & $  5.9 \pm 1.1 $ \\
7  & $12.1 \pm 0.9 $ & $  8.4  \pm 1.2  $ & $ 11.6 \pm 0.9 $ & $ 5.0 \pm 0.7 $ &   \nodata   & $  9.2 \pm 1.4^{1}$ &   \nodata   & $  6.2 \pm 0.9 $ \\
8  &     \nodata     & $ 17.19 \pm 0.17 $ & $ 12.9 \pm 0.5 $ &     \nodata     &   \nodata   & $ 13.2 \pm 1.0 $    &   \nodata   & $  9.5 \pm 1.0 $ \\
9  & $ 8.7 \pm 1.9 $ & $ 13.2  \pm 1.3  $ & $ 14.1 \pm 0.5 $ &     \nodata     &   \nodata   & $ 12.8 \pm 1.0 $    &   \nodata   & $ 10.6 \pm 0.8 $ \\
10 &     \nodata     & $ 15.8  \pm 0.2  $ & $ 13.1 \pm 0.6 $ &     \nodata     &   \nodata   & $ 11.9 \pm 1.6 $    &   \nodata   & $  9.6 \pm 1.4 $ \\
11 &     \nodata     & $ 15.9  \pm 0.2  $ & $ 14.1 \pm 0.8 $ &     \nodata     &   \nodata   & $ 14.8 \pm 0.7 $    &   \nodata   & $  8.6 \pm 1.0 $ \\
\cutinhead{2008 Outburst}
1  &   \nodata   & $ 15.0 \pm 0.2 $ & $ 16.4 \pm 0.5 $ &   \nodata   &   \nodata   & $ 11.8 \pm 1.2 $ &   \nodata   & $ 10.4 \pm  1.7 $ \\
2  &   \nodata   & $ 15.5 \pm 0.3 $ & $ 13.0 \pm 0.7 $ &   \nodata   &   \nodata   & $ 19.1 \pm 0.9 $ &   \nodata   & $  9.1 \pm  1.6 $ \\
3  &   \nodata   & $ 15.4 \pm 0.2 $ & $ 13.6 \pm 0.5 $ &   \nodata   &   \nodata   & $ 15.8 \pm 0.7 $ &   \nodata   & $  9.1 \pm  0.8 $ \\
4  &   \nodata   & $ 15.3 \pm 0.5 $ & $ 14.2 \pm 0.8 $ &   \nodata   &   \nodata   &     \nodata      &   \nodata   & $ 12.3 \pm  1.2 $ \\
5  &   \nodata   &     ---$^2$      & $ 29.2 \pm 1.0 $ &   \nodata   &   \nodata   &     \nodata      &   \nodata   & $ 14.1 \pm  1.6 $ \\
6  &   \nodata   & $ 11   \pm 2   $ & $ 14   \pm 3   $ &   \nodata   &   \nodata   &     \nodata      &   \nodata   &       \nodata     \\
7  &   \nodata   & $ 17.9 \pm 0.8 $ & $ 23.5 \pm 1.1 $ &   \nodata   &   \nodata   &     \nodata      &   \nodata   &       \nodata     \\
\cutinhead{2011 Outburst}
1  &     \nodata     &      ---$^2$     & $ 20.2 \pm 0.9 $ &     \nodata     &   \nodata   & $ 17   \pm 8   $ &     \nodata     & $ 11.5 \pm 0.8 $ \\
2  &     \nodata     &      ---$^2$     & $ 14   \pm 3   $ &     \nodata     &   \nodata   &     \nodata      &     \nodata     & $ 13   \pm 4   $ \\
3  &     \nodata     &      ---$^2$     & $ 23.7 \pm 1.2 $ & $ 7.2 \pm 1.4 $ &   \nodata   & $  7.3 \pm 1.1 $ &     \nodata     & $ 13.4 \pm 0.6 $ \\
4  &     \nodata     &      ---$^2$     & $ 18.5 \pm 1.7 $ & $ 9   \pm 3   $ &   \nodata   & $  9.2 \pm 1.9 $ &     \nodata     & $ 13.4 \pm 0.7 $ \\
5  & $ 6.8 \pm 1.8 $ & $ 14.3 \pm 1.1 $ & $  7.8 \pm 1.4 $ & $ 5.3 \pm 1.2 $ &   \nodata   &     \nodata      & $ 4.1 \pm 0.8 $ & $  7.1 \pm 0.9 $ \\
6  &     \nodata     & $ 17.0 \pm 0.5 $ & $ 12.3 \pm 1.2 $ &     \nodata     &   \nodata   & $ 15.9 \pm 1.7 $ &     \nodata     & $ 11   \pm 2   $ \\
7  &     \nodata     & $ 16.5 \pm 0.6 $ & $ 15.4 \pm 1.2 $ &     \nodata     &   \nodata   &     \nodata      &     \nodata     & $ 21.2 \pm 1.8 $ \\
8  &     \nodata     & $ 16.4 \pm 0.6 $ & $ 21.4 \pm 0.7 $ &     \nodata     &   \nodata   &     \nodata      &     \nodata     &   \nodata   \\
9  &     \nodata     & $ 22.0 \pm 0.8 $ & $ 18.7 \pm 1.2 $ &     \nodata     &   \nodata   &     \nodata      &     \nodata     & $ 28   \pm 3   $ \\
10 &     \nodata     & $ 33.8 \pm 1.1 $ & $ 29   \pm 4   $ &     \nodata     &   \nodata   &     \nodata      &     \nodata     &   \nodata   \\
\enddata
\end{deluxetable*}

\begin{deluxetable}{lccc}
\tabletypesize{\scriptsize}
\tablecaption{ 
    Schechter function fit parameters.
	\label{tab:Schechter} }
\tablehead{
    \colhead{Interval} & \colhead{$\alpha$} & \colhead{$\nu_{\cut}$} & \colhead{r} \\
    \colhead{} & \colhead{} & \colhead{(Hz)} & \colhead{(\%)}
}
\startdata
\cutinhead{2008}
5   & $-0.52  \pm 0.02$  & $3.33 \pm 0.10 $ & $28.4  \pm 0.3 $ \\
\cutinhead{2011}
1   & $-0.450 \pm 0.013$ & $5.63 \pm 0.09 $ & $25.48 \pm 0.10$ \\
2   & $-0.79  \pm 0.15 $ & $5.6  \pm 1.0  $ & $15.4  \pm 1.3 $ \\
3   & $-0.59  \pm 0.02$  & $4.95 \pm 0.17 $ & $26.75 \pm 0.15$ \\
4   & $-0.60  \pm 0.02$  & $5.67 \pm 0.16 $ & $27.1  \pm 0.3 $ \\
\enddata
\end{deluxetable}

	\subsection{1998}
        The first \textit{RXTE} detection of SAX~J1808 was with a scan observation on 
        MJD~50915 \citep[day~9 of Figure~\ref{fig:OutburstEvolution},][]{IAUC.6876}, 
        during which the source flux was $59$~mCrab.  At this time the source 
        was in the island state showing a power spectrum that can be fitted with 
        three Lorentzians ($L_b$, $L_h$
        and $L_u$).
        
        The first pointed \textit{RXTE} observation was five days later on MJD~50920,
        at which time the flux had decayed to $\sim$43~mCrab. The soft color
        was decreasing, while the hard color increased.
        The power spectrum shifted to lower frequencies and requires an
        additional $L_{\low}$ component, clearly showing that SAX~J1808
        transitioned to the extreme island state.  

        Over the following 14~days the flux and soft colors steadily decreased,
        while the hard color remained constant for 5 days before also starting
        to decay. The flux and soft color evolution transitioned into a faster 
        decay on MJD~50929 (day~23). The
        flux decay briefly slowed down three days later. Finally the flux
        dropped below the detection limit on MJD~50934 (day~28), as the source
        returned to quiescence. During the entire decay SAX~J1808 remained in
        the extreme island state. 
		
	\subsection{2002}
        In the 2002 outburst SAX~J1808 was first detected with \textit{RXTE} on MJD
        52560. When pointed \textit{RXTE} observations commenced on MJD~52562
        (day~6) the source was already close to the flux maximum.
        In interval~1 the power spectrum clearly reflects the island state, showing 
        $L_b$, $L_h$, $L_{\hHz}$ and $L_u$.
        In interval~2 both colors decrease as SAX~J1808 appears to be
        transitioning to the lower-left banana. The power spectrum now shows a
        complex morphology, requiring at least six Lorentzian components.  At
        $\sim$50~Hz the $L_{\LF}$ component appears, and we detect a $2.8\sigma$
        feature at $\nu_{0}=434\pm12$~Hz which we tentatively accept as the lower
        kHz QPO, $L_\ell$, as the upper kHz QPO is seen simultaneously at $\nu_{0,u} = 597.9\pm5.3$~Hz. 

        In the following observations a flux maximum of $\sim$85~mCrab is reached, and
        maintained for about a day. The lower kHz QPO is detected at
        higher significance with $\nu_{0,\ell} = 503.5\pm3.1$~Hz and $\nu_{0,u} = 684.5\pm4.9$~Hz, 
        and a second noise component appears near $L_b$,
        providing further evidence that the source indeed transitioned to the
        lower-left banana. 
		
        As the flux starts to decay on MJD~52564 (day~8), both colors increase
        and SAX~J1808 transitions back into the island state. From interval~7
        (day~9) the soft color starts to decrease once more, while the hard
        color remains roughly constant.  The power spectrum shows very little
        change and remains consistent with an island state morphology. 
		
        After reaching a flux minimum on MJD~52579 (day~23), SAX~J1808 entered
        the prolonged outburst tail \citep{Patruno2009}, which continued for 
        24~days.
	
	\subsection{2005}
	\label{sec:2005}
        Routine monitoring with \textit{RXTE} revealed a new outburst of SAX~J1808
        on MJD~53521 (day~1) \citep{Markwardt2005}. In the following days the
        flux steadily increased until a maximum of $\sim$60~mCrab was reached
        approximately five days after first detection. 
        
        While the flux increased to its maximum (intervals~1--6) the hard color showed a
        steady decay, whereas the soft color remained roughly constant.
        During this time the power spectrum is that of the island state, 
        showing $L_b$, $L_h$, $L_{\hHz}$ and $L_u$. 
        
        The flux maximum of 60~mCrab is reached on MJD~53528 (day~6, interval~7).
        In this interval both colors drop sufficiently to suggest a transition
        to the lower-left banana. The power spectrum confirms this transition
        with the appearance of $L_{b2}$ and a narrow QPO at 58~Hz, which
        we identify as $L_{\LF}$. Additionally a broad feature appears at
        $\nu_{\max}=428$~Hz, which is difficult to
        identify as the low $Q$ suggests it is $L_{\hHz}$, while the frequency suggests
        it is $L_{\ell}$. The amplitude of 9\% rms lies roughly in between
        $L_{\hHz}$ and $L_{\ell}$. We label it as $L_{\hHz}$, noting that it 
        is not a `clean' detection.
        
        After reaching the flux maximum, both colors increased as SAX~J1808
        transitioned to the island state. Over the following seven days the flux
        and soft color decayed, while the hard color increased.  On MJD~53542
        (day 20) SAX~J1808 was only marginally detected, marking the end of the
        main outburst and the start of a 40~day prolonged outburst tail
        \citep{Patruno2009}.
		
	\subsection{2008}
        On MJD~54731 (day 1) SAX~J1808 was again detected in outburst with
        \textit{RXTE}. As in 2005 the rise to maximum flux was well sampled, and
        showed the same behavior: as the flux increased, the hard color
        decreased and the soft color remained constant. The power
        spectrum is an island state one and is well described with four
        Lorentzians ($L_b$, $L_h$, $L_{\hHz}$ and $L_u$.)
        
        During its luminosity maximum (intervals~3--4, days~3--7) SAX~J1808
        reached a flux of $55$~mCrab, the lowest flux maximum of all outbursts.
        The soft and hard colors again show a drop prior to the onset of the
        flux decay. Although this drop is as deep as seen in the other
        outbursts, the power spectrum indicates the source does not make a state
        transition to the lower-left banana. A sharp feature does appear at
        $\nu=283$~Hz. Given $\nu_u=520$~Hz this is roughly where $L_{\ell}$ is
        expected to appear, but at 2$\sigma$ it is not significant.
		
        On MJD~54737 (day~6) both colors increase as the flux starts to decay.
        At this point (interval~5) the power spectrum shows an unexpected
        deviation from the regular island state morphology as the high
        luminosity flaring now dominates the power spectrum below 10~Hz
        \citep{Bult2014}. With a fractional rms amplitude of $\sim$30\%, the
        flaring overwhelms the $L_b$ component that is normally seen in the
        same frequency range. We therefore cannot fit the $L_b$ component to
        the data. The flaring remains present for about three days, after which
        the power spectrum returns to the regular island state shape. 

        On MJD~54748 (day~19) SAX~J1808 enters the prolonged outburst tail,
        which lasts for 30~days.
	
	\subsection{2011}
        The first pointed \textit{RXTE} observation of SAX~J1808's seventh outburst was
        on MJD~55869 (day~2), roughly five days after the initial detection with
        \textit{SWIFT} \citep{Markwardt2011}. The low frequency region of the power
        spectrum was again dominated by high luminosity flaring. In the first
        two intervals the mid/high frequency domain required three Lorentzians
        ($L_h$, $L_{\hHz}$ and $L_u$), which suggests that despite the low color
        values the source was still in the island state. 
        
        On MJD~55871 (day~4, interval~3), a maximum flux of 80~mCrab was reached,
        which correlated with a drop in amplitude of the flaring component. In
        the next interval the amplitude of the flaring increased again and the
        high frequency components suggested the source transitioned to the
        lower-left banana. The feature which is normally identified as the hump
        component has a central peak, but also broad wings and hence is not well described
        by a single Lorentzian profile, but instead requires two Lorentzians for
        a satisfactory fit: one to describe the sharp QPO and one for the overlapping
        noise component. We identify the broader noise component as $L_{h}$
        and the peaked QPO as $L_{\LF}$, but note that this identification is
        uncertain as the two components are not well resolved, and may be distorted
        by the high frequency wing of the high luminosity flaring component.
        
        In interval~5 (day~7) the high luminosity flaring disappears, while colors
        remain at low values. The $L_{\LF}$ component is now clearly present in the power
        spectrum and the marginal detection ($2.8\sigma$) of the lower kHz QPO at $\nu_{0,\ell} =
        566.6\pm9.4$~Hz with a simultaneous $\nu_{0,u} = 736.1\pm9.7$~Hz confirms SAX~J1808 entered the lower-left banana.
        
        In interval~6 (day~8) the colors increased as the source switched back to the
        island state. The power spectrum can be described with $L_b$,
        $L_h$, $L_{\hHz}$ and $L_u$.
        In the following ten days SAX~J1808 showed a regular decay with the soft
        color decreasing with flux while the hard color increased. During this
        time the power spectral components smoothly move to lower frequencies, before
        transitioning into the extreme island state in the last two intervals.
	
	\subsection{The 410 Hz QPO}
	\label{sec:410results}
        In the outburst of 2002, between MJD~52565 and MJD~52573, \citet{Wijnands2003} 
        detected a sharp QPO at $\sim$410~Hz with an amplitude of $\sim$2.5\%~rms in 
        the 3--60~keV energy range. We search all outbursts for this 410~Hz QPO using 
        our 2--20~keV band, which gives a better signal
        to noise ratio. Due to the small amplitude of the QPO, we nearly always 
        have to average multiple intervals to obtain a significant detection. 
        
        We detect the 410~Hz QPO at $>3\sigma$ significance in the outbursts of 
        2002, 2005 and 2008 (see Table~\ref{tab:410HzParameters}), but not in 
        the outbursts of 1998 and 2011, for which we obtain 95\% confidence upper 
        limits for the amplitude of $\sim$3\%. The QPO amplitude varies between 2.1\% and
        3.6\%, and shows no relation with $\nu_{410}$ or $\nu_{u}$. The
        characteristic frequency varies smoothly between $409.6$ and $416.6$~Hz 
        and is correlated with the upper kHz QPO frequency. 
		
\begin{deluxetable*}{lccccccc}
\tabletypesize{\scriptsize}
\tablecaption{ 
    410~Hz QPO Fit Parameters
	\label{tab:410HzParameters} }
\tablehead{ 
    \colhead{Intervals} & \colhead{$\nu_{410}$} & \colhead{Q} & \colhead{r} & \colhead{P/$\sigma_P$}
        & \colhead{$\nu_{b}$} & \colhead{$\nu_{h}$} & \colhead{$\nu_{u}$} \\
    \colhead{ } & \colhead{(Hz)} & \colhead{} & \colhead{(\%)} & \colhead{ }
        & \colhead{(Hz)} & \colhead{(Hz)} & \colhead{(Hz)}
}
\startdata
\cutinhead{2002}
6, 7, 8      & $409.6 \pm 0.6$  & $ 80^{+29}_{-21}$  &  $2.1\pm0.3$  &  4.4  & $3.07 \pm 0.04$ & $18.30 \pm 0.10$ & $330 \pm  3$  \\
10, 11       & $414.7 \pm 0.6$  & $ 70^{+28}_{-17}$  &  $3.6\pm0.5$  &  4.3  & $4.95 \pm 0.11$ & $25.3  \pm 0.3 $ & $383 \pm  6$  \\
12, 13, 15   & $410.8 \pm 1.1$  & $ 41^{+17}_{-12}$  &  $3.4\pm0.5$  &  3.6  & $4.02 \pm 0.07$ & $22.1  \pm 0.2 $ & $349 \pm  7$  \\
\cutinhead{2005}                                                                                                                  
3            & $416.6 \pm 0.8$  & $ 98^{+49}_{-21}$  &  $3.2\pm0.5$  &  4.1  & $5.51 \pm 0.09$ & $28.2  \pm 0.3 $ & $390 \pm  7$  \\
\cutinhead{2008}                                                                                                                  
1, 2, 3      & $416.5 \pm 0.9$  & $ 92^{+92}_{-38}$  &  $2.7\pm0.5$  &  3.4  & $4.55 \pm 0.08$ & $22.8  \pm 0.4 $ & $399 \pm 10$  \\
\enddata
\tablecomments{
    Fit parameters for the 410 Hz feature, its significance ($P/\sigma_P$), and
    the simultaneously measured $\nu_{\max}$ of break and hump components and 
    the upper kHz QPO.
}
\end{deluxetable*}

\begin{figure}
			\includegraphics[width=1.0\linewidth]{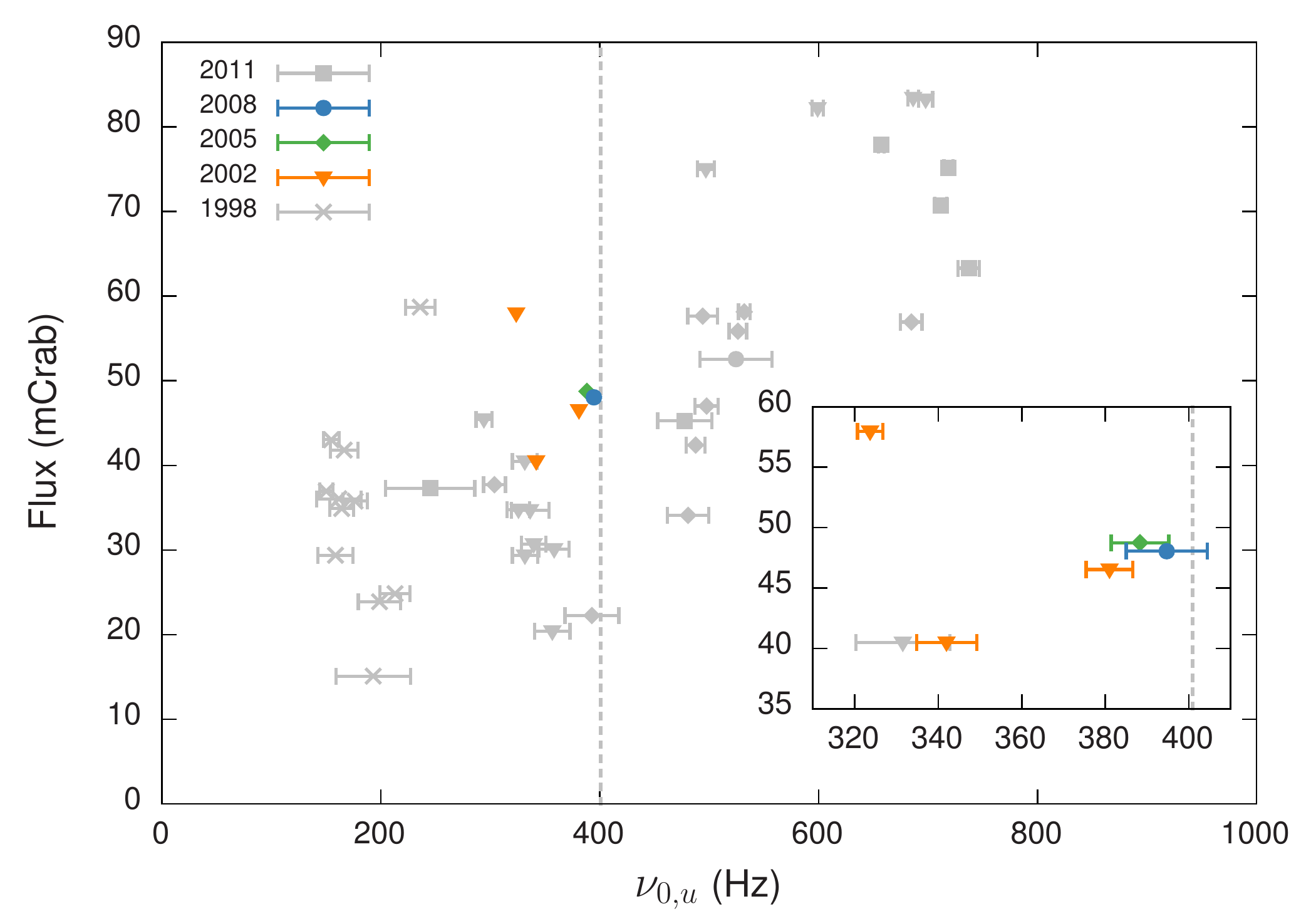}
			\caption{ 
				Overview of the 410~Hz QPO detections. We show flux vs.
				the upper kHz QPO centroid frequency for all intervals in
				grey. Intervals in which the 410~Hz QPO was detected are marked
				in yellow (2002), green (2005), and blue (2008). The vertical dashed line
				marks the spin frequency ($\nu_{\text{spin}}=401$~Hz). The inset shows a
				zoom-in of the detection region.
			}
			\label{fig:410HzOverview}
		\end{figure}

        We find the 410~Hz QPO is present in all observations in which
        $L_u$ has $320 \leq \nu_{0} \leq 401$~Hz and the flux is higher than
        40~mCrab (Figure~\ref{fig:410HzOverview}). An exception is interval~14 
        of the 2002 outburst (grey point in Figure~\ref{fig:410HzOverview} zoom-in), for which we
        find a 95\% confidence upper limit on $r_{410}$ of 3.4\%; comparable to the
        typical amplitude. Due to their lower count rates, the observations below
        40~mCrab have higher upper limits on the amplitude ($\sim$4\%) and are 
        therefore also consistent with the QPO being below the detection threshold (also see 
        section~\ref{sec:410discussion}).
		
        The 410~Hz QPO is only seen in the island state, but is not related to
        a specific stage of the outburst evolution. In 2005 the QPO was
        detected during the rise to outburst, in 2008 at the maximum luminosity,
        while in 2002 the QPO was seen during the flux decay.   
		
		The proximity to the 401~Hz spin frequency suggests $L_{410}$ may be an upper
		sideband of the pulsations. If this is the case then a similar feature might
		be present at the lower sideband frequency of $\sim$392~Hz. Using the same data selection
		for which we detected $L_{410}$, we searched for the lower sideband, but no significant 
		features were found, giving a 95\% confidence upper limit on the fractional rms of 
		1.1\% to 2.3\%. 
		To improve sensitivity we use the shift-and-add method \citep{Mendez1998} in
		which we shift the power spectra before averaging them, such that the frequency of the predicted
		lower sideband feature, $2\nu_{\text{spin}}-\nu_{410}$, is expected at
		390~Hz. Again, a lower sideband is not observed.

\section{Discussion}
    It is common practice to study the variability properties of X-ray binaries
    by considering the relation between the frequencies of several components
    of the power spectrum. This approach is based on the success of the WK
    \citep{Wijnands1999} and PBK \citep{Psaltis1999} relations, which show
    frequency correlations over several orders of magnitude, linking the
    variability components of neutron star and black hole binaries.
	
    Applied to atoll sources the WK relation considers the correlation between the break frequency 
    versus the $L_{\LF}$ or $L_{h}$ frequency, whereas the PBK relation 
    considers the correlation between the low frequency QPOs
    and the second highest frequency feature in the power spectrum. For the
    higher luminosity lower-left banana state this would be $L_{\ell}$, while 
    for the lower luminosity extreme island state it is $L_{\low}$. In the island
    state it is unclear which component should be considered as neither $L_{\low}$ or
    $L_{\ell}$ are observed (also see section~\ref{sec:ell-vs-low}).
	
    For atoll sources \citet{Straaten2003} suggested a scheme of correlations
    for the frequencies of several Lorentzian components plotted against the
    frequency of $L_u$. This scheme of correlations encompasses the WK and PBK
    relations and is the same across different atoll sources. AMXPs also follow this
    atoll correlation scheme, although they tend to be shifted in their $\nu_u$
    by a constant factor of 1.1--1.6, which differs per source
    \citep{Straaten2005, Linares2005}.
	
    A complication in considering frequency relations comes from the fact that the 
    characteristic frequency of the Lorentzian function may not be the 
    physical frequency of the variability mechanism. In this work we used $\nu_{\max}$ as the
    characteristic frequency, which allows to treat QPO and noise components on the same 
    footing, and tends to reduce the scatter on frequency correlations \citep{Belloni2002}. 
    However, models involving resonances or beat frequencies tend to be defined in terms of
    the centroid frequency $\nu_0 = \nu_{\max}/\sqrt{1+1/(4Q^2)}$. For narrow QPOs (high Q)
    these two frequencies are almost the same, while for broad noise components (low Q)
    $\nu_{0}$ is significantly smaller than $\nu_{\rm max}$. 
    We will continue to consider the frequency relations in terms of $\nu_{\max}$ unless 
    explicitly stated otherwise.
    
    Further difficulty in interpreting frequency relations comes from
    ambiguities in the identification of the variability components. In
    particular when measurements deviate from a correlation, it could
    either indicate a break down of the relation, or a mislabeling of
    the considered component. It is therefore useful to distinguish between
    secure and more uncertain identification of the power spectral components.
    
	\begin{figure*}[p]
		\includegraphics[width=1.0\linewidth]{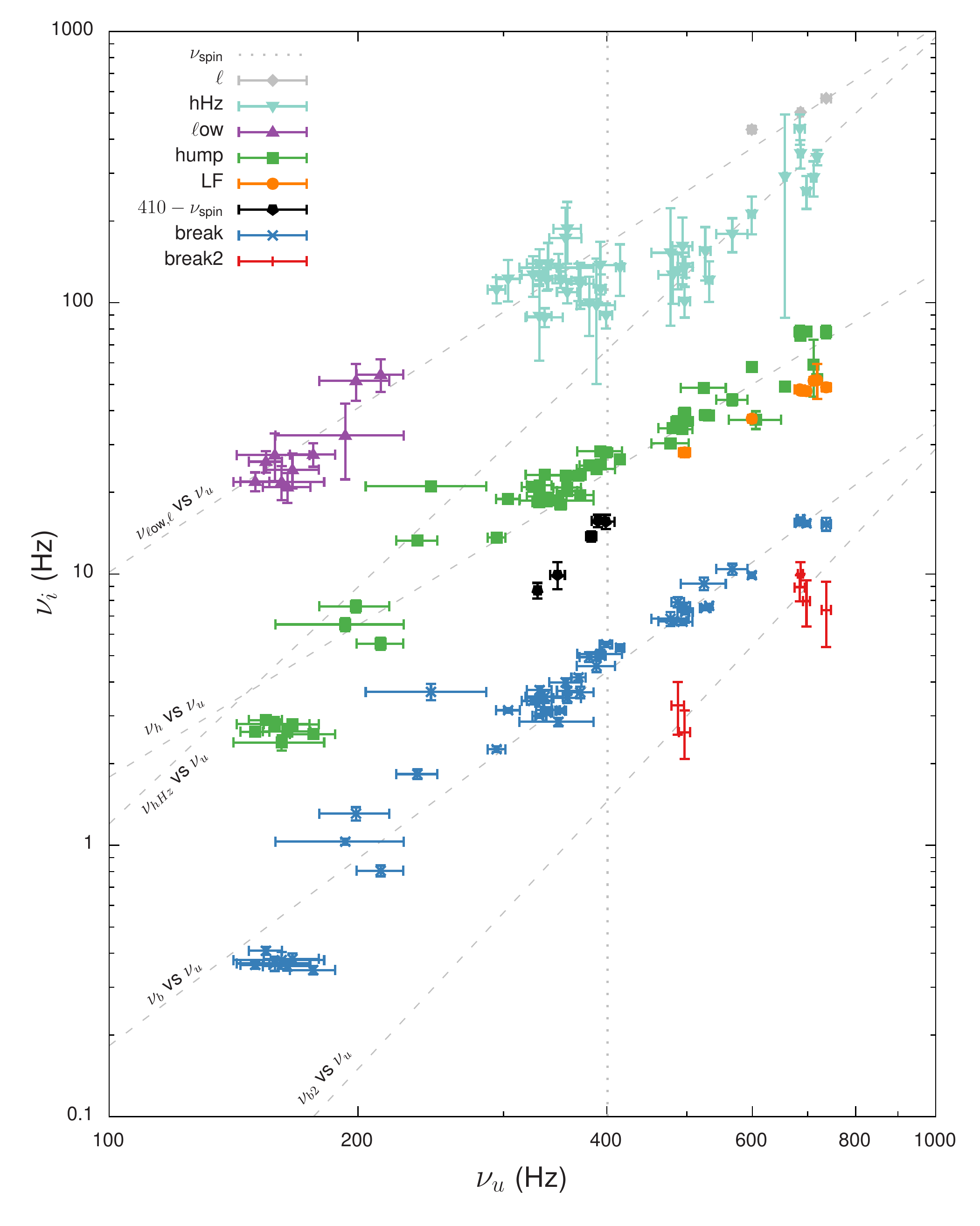}
		\caption{
			Characteristic frequencies of the power spectral components
			as a function of $\nu_u$. Each point corresponds to a frequency
			measurement in Table~\ref{tab:Frequencies}. The dashed lines
			show the fitted power-law relations (see Table~\ref{tab:PowerLaws}) 
			and the vertical dotted line gives the frequency of the 401~Hz pulsations.				
			For clarity we omitted points for which $Q_u$ was fixed
			at zero, as these points have large errors in $\nu_u$. 
		}
		\label{fig:AtollScheme}
	\end{figure*}

\begin{deluxetable*}{llccc}
\tabletypesize{\scriptsize}
\tablecaption{ 
    Power-law Fit Parameters
	\label{tab:PowerLaws} }
\tablehead{
    \colhead{Component} & \colhead{Data selection$^\text{a}$} & \colhead{Normalization} 
    & \colhead{Index} & \colhead{$\chi^2/dof$}
}
\startdata
\cutinhead{Atoll Scheme Relations ($\nu_i$ vs $\nu_u$)}
break	& all					& $ (4.8 \pm 0.7)\times10^{-6}  $ & $ 2.29 \pm 0.02 $ & 225/47 \\
hump	& all     				& $ (3.4 \pm 0.4)\times10^{-4}  $ & $ 1.86 \pm 0.02 $ & 748/51 \\
hHz	 	& $\nu_u>500\ {\rm Hz}$ & $ (2   \pm 5  )\times10^{-6}  $ & $ 2.9  \pm 0.4  $ &   6/7  \\
LF	 	& all     				& $ (1.2 \pm 1.0)\times10^{-3}  $ & $ 1.62 \pm 0.12 $ & 6.1/6  \\
b2          & all  				& $ (5   \pm14  )\times10^{-9}  $ & $ 3.3  \pm 0.5  $ & 5.3/4  \\
$\low$      & all 				& $ (1   \pm 3  )\times10^{-6}  $ & $ 3.4  \pm 0.8  $ &   6/9  \\
$\low$ \& $\ell$  & all			& $(10   \pm 2  )\times10^{-4}  $ & $ 2.01 \pm 0.04 $ &  28/12 \\
LF \& ($410-401$) & all			& $ (1.2 \pm 0.9)\times10^{-4}  $ & $ 1.97 \pm 0.11 $ &  58/11 \\
\cutinhead{State Selected Relations ($\nu_i$ vs $\nu_u$)}
break \& b2 & EIS \& all$^\text{b} $ & $ (7   \pm 2  )\times10^{-6}  $ & $ 2.17 \pm 0.05 $ &  129/15 \\
break       & IS                     & $ (2.1 \pm 0.6)\times10^{-5}  $ & $ 2.05 \pm 0.05 $ &  110/32 \\
hump        & IS                     & $ (1.4 \pm 0.2)\times10^{-3}  $ & $ 1.64 \pm 0.02 $ &  519/35 \\ 
break       & EIS                    & $ (0.4 \pm 6.6)\times10^{-19} $ & $ 8.6  \pm 2.6  $ &   19/9 \\
hump        & EIS                    & $ (0.2 \pm 3.5)\times10^{-17} $ & $ 8.2  \pm 2.5  $ &   19/9 \\
\cutinhead{WK relations ($\nu_i$ vs $\nu_b$)}
break2		& all					& $(10   \pm 6)\times10^{-2}    $ & $ 1.7  \pm 0.2  $ & 1.3/4  \\
LF			& all					& $  7.9 \pm 1.0				$ & $ 0.66 \pm 0.05 $ & 6.8/4  \\
hump		& all					& $  6.6 \pm 0.1  				$ & $ 0.88 \pm 0.01 $ & 186/54 \\
$\low$		& all					& $ 44   \pm 3   				$ & $ 0.60 \pm 0.09 $ &  15/9  \\
u 			& all					& $208   \pm 2  				$ & $ 0.44 \pm 0.01 $ & 231/49 \\
LF \& ($410-401$) & all				& $  3.5 \pm 0.7                $ & $ 0.96 \pm 0.07 $ &  83/9  \\
\cutinhead{PBK relations ($\nu_i$ vs $\nu_{\low/\ell}$)}
break			& all & $(1.2 \pm 0.1)\E{-2}$ & $1.15 \pm 0.01$ & 215/13 \\
hump			& all & $(1.12\pm0.01)\E{-1}$ & $1.04 \pm 0.01$ & 148/12 \\
upper kHz QPO	& all & $31.2\pm1.8$          & $0.50 \pm 0.01$ &  29/13 \\
\enddata
\tablecomments{Results of power-law fits to the frequency-frequency relations.}
\tablenotetext{a}{IS = Island State, EIS = Extreme Island State}
\tablenotetext{b}{i.e. EIS data for the break component and all data of the break2 component.}
\end{deluxetable*}

	\subsection{Frequency Correlations}
	\label{sec:correlations}
        Figure~\ref{fig:AtollScheme} shows the frequency relations with respect
        to $\nu_u$ for all outbursts of SAX~J1808. Overall, the measured
        frequencies follow the trends of the atoll correlation scheme. We
        quantify the relations by fitting a set of power-laws over the full
        $\nu_u$ frequency range and present the fit parameters in
        Table~\ref{tab:PowerLaws}. We now discuss the notable features in this
        figure.
	
		\subsubsection{Break and Hump}
		\label{sec:break-hump}
			The break and hump components are well correlated with $\nu_u$ (see 
			Figure~\ref{fig:AtollScheme}), but that correlation shows a break at 
			$\nu_u\simeq250$~Hz. This break could be due to a misidentification, 
			however, as we are considering the three dominantly present power 
			spectral components this is unlikely.
			
\begin{figure*}[p]
				\includegraphics[width=\linewidth]{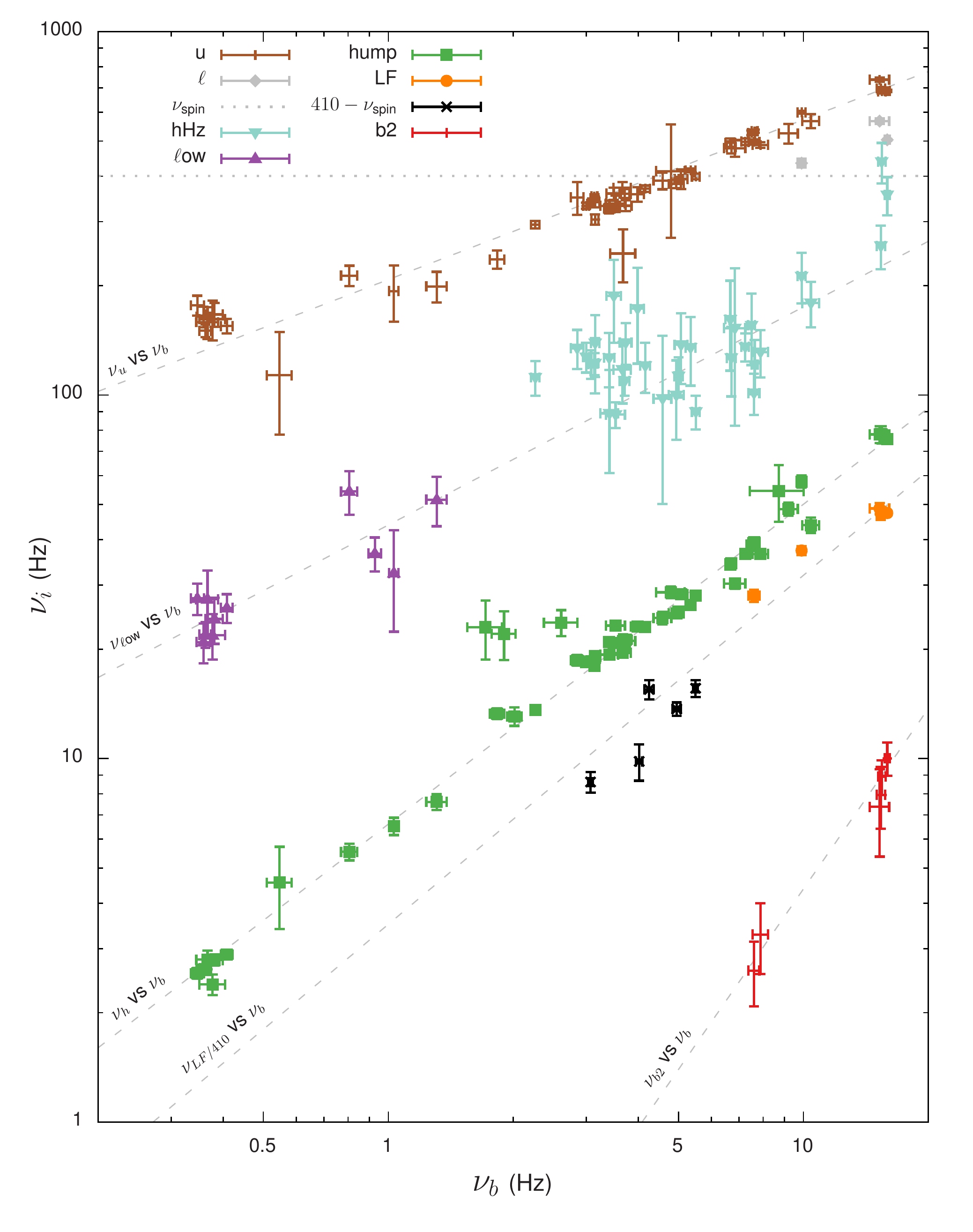}
				\caption{ 
					Characteristic frequencies of the power spectral components as a function
					of $\nu_{b}$. The green squares show the hump vs break frequency (WK relation).
					For further details see Figure~\ref{fig:AtollScheme}.
				}
				\label{fig:WK-Basic}
			\end{figure*}

			The break in the correlation occurs at the transition between the island 
			state and the extreme island state and appears to affect the break and hump
			components in exactly the same way. Indeed, if we consider the hump 
			frequency as a function of the break frequency (i.e. the WK relation) we find the 
			two components are tightly correlated (Figure~\ref{fig:WK-Basic}) and do not
			show state dependent behavior. This suggests that the break seen in 
			Figure~\ref{fig:AtollScheme} is due to a change in $\nu_u$ that occurs when the source
			transitions between the island and extreme island states. 
			
			When considering centroid frequencies instead of characteristic frequencies,
			the break remains present. Additionally, for centroid frequencies the WK 
			relation no longer holds, with $L_{h}$ drifting in frequency faster than $L_{b}$.
			
			If the upper kHz QPO indeed shows state dependent variations that do not occur in the
			other components, then considering frequency correlations against $\nu_u$ is not optimal. 
			In Figure~\ref{fig:WK-Basic}
			we therefore plot the frequencies of all components agains $\nu_b$ (similar to 
			the $\nu_{\rm band}$ approach of \citealt{Straaten2002}), which does not show this
			state dependent variation. 

		\subsubsection{$L_{\low}$ and $L_\ell$}
		\label{sec:ell-vs-low}
			The relation between $L_{\low}$ and $L_{\ell}$ is a long standing problem.
			The two features were proposed to be due to the same component by \citet{Psaltis1999},
			who suggested that like the break and hump components, $L_{\low}$ too becomes increasingly
			coherent for higher frequencies, eventually evolving into $L_{\ell}$. 
			Plotting the frequency of $L_h$ against that of $L_{\low}$ or $L_{\ell}$ then reveals
			a tight correlation which can be extended over three orders of magnitude by
			linking black hole and neutron star data.

			The uncertainty of the PBK relation comes from the fact that the transition from 
			$L_{\low}$ to $L_{\ell}$ has never been observed. It has also been suggested
			that the two features are actually distinct components \citep{Straaten2003,
			Straaten2005}. For SAX~J1808 specifically \citet{Straaten2005} suggested
			the two components were not the same because each requires a different frequency
			shift to be compatible with the atoll correlation scheme. 
			
\begin{figure}
				\includegraphics[width=\linewidth]{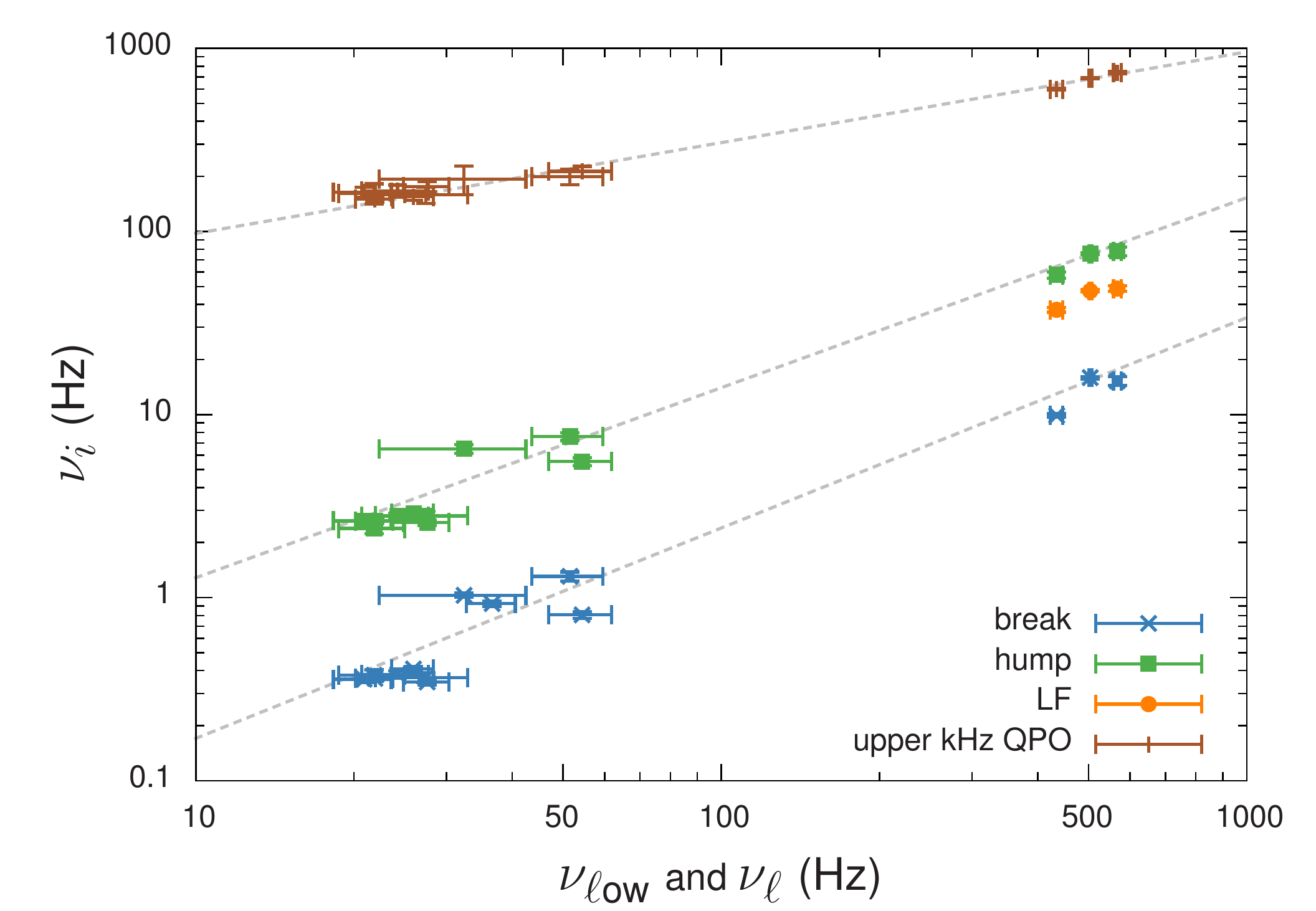}
				\caption{
					Frequency measurements plotted against the $L_{\low}$ ($<100$~Hz)
					and $L_{\ell}$ ($>100$~Hz) characteristic frequencies (PBK relation). Dashed lines show
					fitted power-law relations (see Table~\ref{tab:PowerLaws}). \\
				}
				\label{fig:PBK}
			\end{figure}

			Since most observations of SAX~J1808 are in the island state, they do not have a
			$L_{\low}$ or $L_{\ell}$ measurement. For those observations that do have either component
			present we plot the measured frequencies against $\nu_{\low}$ and $\nu_{\ell}$ in 
			Figure~\ref{fig:PBK}. The break and hump components and the upper 
			kHz QPO all follow a correlation with $\nu_{\low/\ell}$. Corresponding to the 
			break below $\nu_u \sim 250$~Hz observed in Figure~\ref{fig:AtollScheme}, the 
			extreme island state $L_u$ appears to have a relatively high frequency, resulting 
			in a flatter correlation for $L_u$ as compared to $L_b$ and $L_h$. The very similar 
			correlations for $L_b$ and $L_h$ are in accordance with the idea that the 
			$L_{\low}$ and $L_{\ell}$ features are the same component, as assumed by PBK, and
			that it is $L_u$ that deviates at low frequency. 
			
			Based on the PBK relation (Figure~\ref{fig:PBK}) $L_{\low}$ and $L_{\ell}$ could
			be the same, however, when considering the frequencies against $\nu_{b}$ 
			(Figure~\ref{fig:WK-Basic}) it seems more likely that $L_{\low}$ is somehow
			connected to $L_{\hHz}$. Whether or not the $\low$ component and the lower kHz QPO
			are the same or different is therefore not clear from the SAX~J1808 data alone.
			
		\subsubsection{The hHz Component}
		\label{sec.low-vs-hhz}
			The hHz component that appears in the power spectra of atoll
            sources is usually seen in a frequency range of 100--200~Hz and does
            not correlate with $\nu_{u}$ \citep{Ford1998}. Yet in SAX~J1808 it
            appears that this hectohertz component does depend on $\nu_u$ 
            (Figure~\ref{fig:AtollScheme}). 
            
            For $\nu_u\lesssim550$~Hz $L_{\hHz}$ appears in the same way as it does
            in other atoll sources, showing a large scatter between 100 and 200~Hz,
            but remaining constant with $\nu_u$ \citep{Straaten2002,Straaten2005}. Yet from $\nu_u\simeq550$~Hz the
            frequency of the hectohertz component starts to systematically increase
            with the upper kHz QPO frequency, in a way that is very similar to the
            trends seen in the other components. 
            
            Systematic trends of $\nu_{\hHz}$ have previously been reported by 
            \citet{Altamirano2008} who see a non-monotonic dependence of $\nu_{\hHz}$
            on $\nu_u$. These variations appear similarly for $\nu_u\gtrsim550$~Hz, yet
            are always within the scatter of $\nu_{\hHz}$ seen at lower $\nu_u$. In
            SAX~J1808, however, the correlation extends well beyond the initial $\nu_{\hHz}$
            scatter, suggesting the trend cannot be attributed to random variations in the 
            frequencies.
            
            It should be noted that $L_{\hHz}$ is not a dominant feature in the power
            spectrum. Its properties may be influenced by the position and amplitude of
            the nearby kHz QPOs or potentially by an unresolved $\low$ component that 
            could exist at similar frequencies.

            The correlation between $L_{\hHz}$ and $L_u$ was already tentatively
            observed by \citet{Straaten2005}, who also suggest it may be related to 
            $L_{\low}$. Indeed if the $\low$ component follows the same frequency 
            trend with $\nu_u$ as the break and hump components, then the island state
            observations should show a blend of $L_{\low}$ and $L_{\hHz}$. Then, as
            $\nu_u$ increases, the $\low$ component could become more prominent, causing
            the blended feature to follow the apparent correlation with $\nu_u$.
            
            If this interpretation is correct, it has a direct consequence for the relation
            between $L_{\low}$ and $L_{\ell}$. In one observation, interval 4 of 2002, the 
            supposed $\low/\hHz$ blend is seen simultaneously with a lower kHz QPO, which
            would imply that $L_{\low}$ cannot be the same
            component as $L_{\ell}$. On the other hand, if $L_{\low}$ and $L_{\ell}$ are in
            fact the same component, like the PBK relation suggests, then the
            observed correlation between $\nu_{\hHz}$ and $\nu_u$ might be intrinsic
            to the hectohertz component.
            Which of these interpretations is correct remains undecided.
                     
		\subsubsection{kHz QPOs}
		\label{sec:kHzQPOs}
			So far, twin kHz QPOs had only been reported once in SAX~J1808 
			\citep{Wijnands2003}, from data corresponding to our interval 4 of the 2002
			outburst. This frequency pair is separated by roughly half the spin
			frequency and is consistent with $\nu_{0,u}=|3\nu_{0,\ell}-2\nu_{\text{spin}}|$
			\citep{Wijnands2003} and a resonance at a 7:5 ratio \citep{Kluzniak2004}.
			
			Unlike the twin kHz QPOs in most atoll sources \citep{Klis2006}, for which
			the lower kHz QPO is the most prominent feature, in SAX~J1808 it is the upper
			kHz that is most prominent. This phenomenology is also seen in the AMXP XTE~J1807--294
			\citep{Linares2008}.

		  	\begin{figure}
				{ \includegraphics[width=\linewidth]{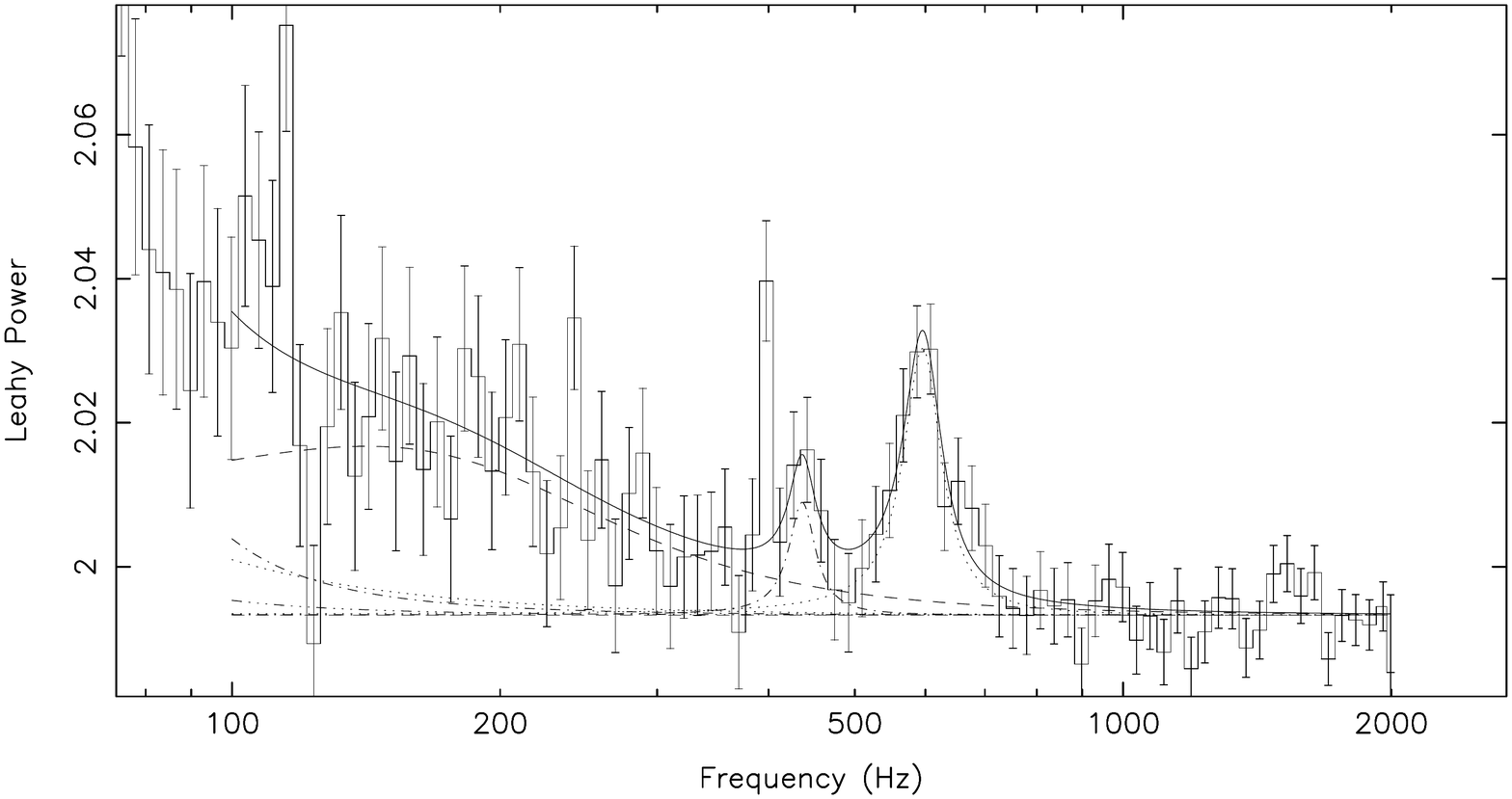} } \vspace{-1.5cm} \\
				{ \includegraphics[width=\linewidth]{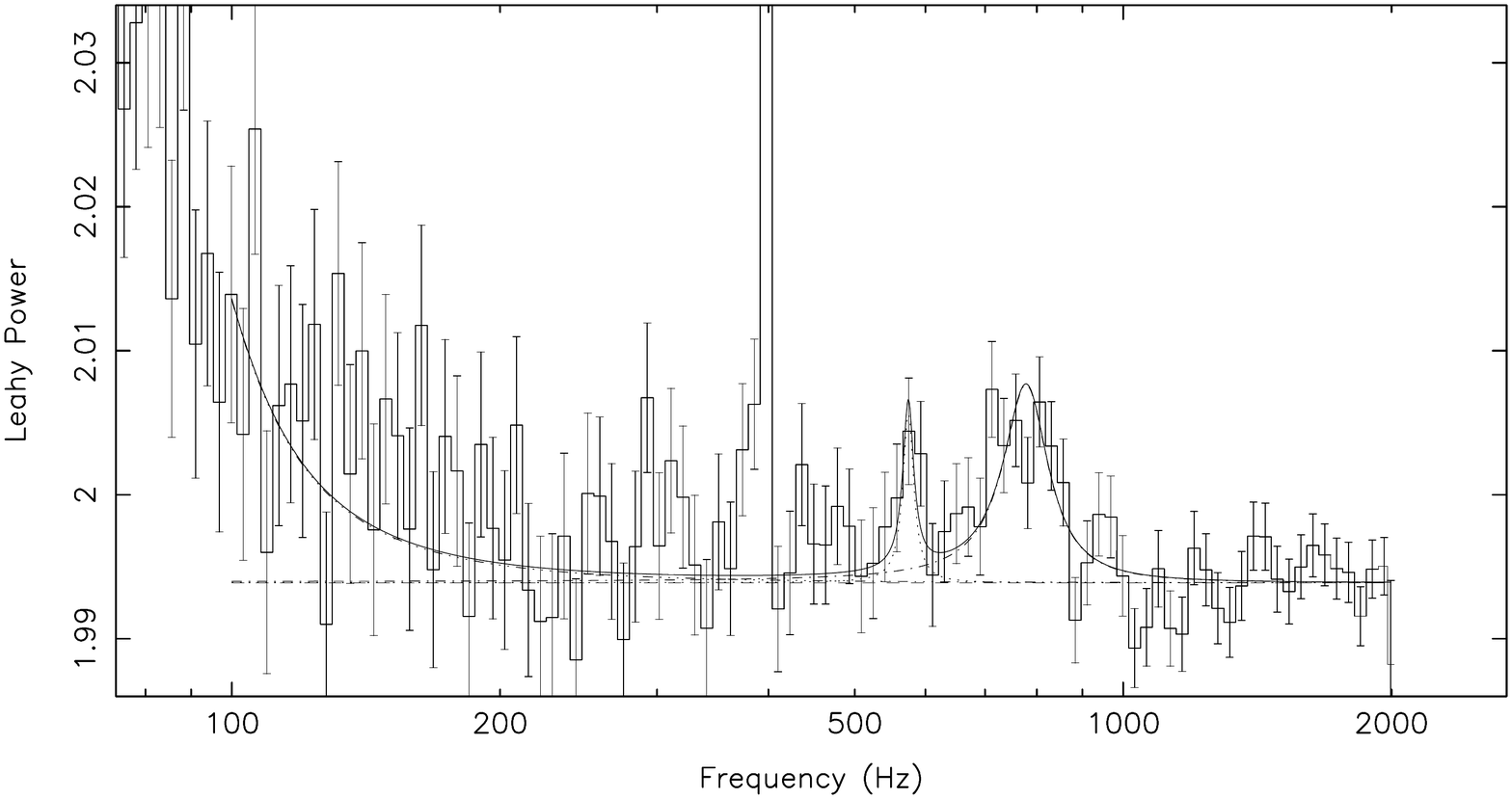} }
				\caption{
					Twin kHz QPO's in SAX~J1808. The top panel shows the twin kHz QPOs in
					interval 2 of the 2002 outburst and the bottom panel those of interval
					5 in the 2011 outburst. The pulse spike at 401~Hz was excluded from
					the fit.
				}
				\label{fig:kHzQPOPDS}
			\end{figure}

			In this work we present additional detections of the twin kHz QPOs in SAX~J1808 in
			interval 2 of the 2002 outburst and interval 5 of the 2011 outburst 
			(Figure~\ref{fig:kHzQPOPDS}). 
			The new detections of $L_{\ell}$ are only marginally significant ($2.8\sigma$),
			however, their phenomenology is entirely consistent with the original detection by
			\citet{Wijnands2003}, providing additional evidence that we indeed observe the twin kHz
			QPO. All three observations are near the maximum luminosity of
			their respective outbursts, at a time when SAX~J1808 has transitioned
			to the lower-left banana. The lower kHz QPO is very narrow and has a lower
			fractional rms amplitude than the upper kHz QPO. 
			
\begin{figure}
				\includegraphics[width=\linewidth]{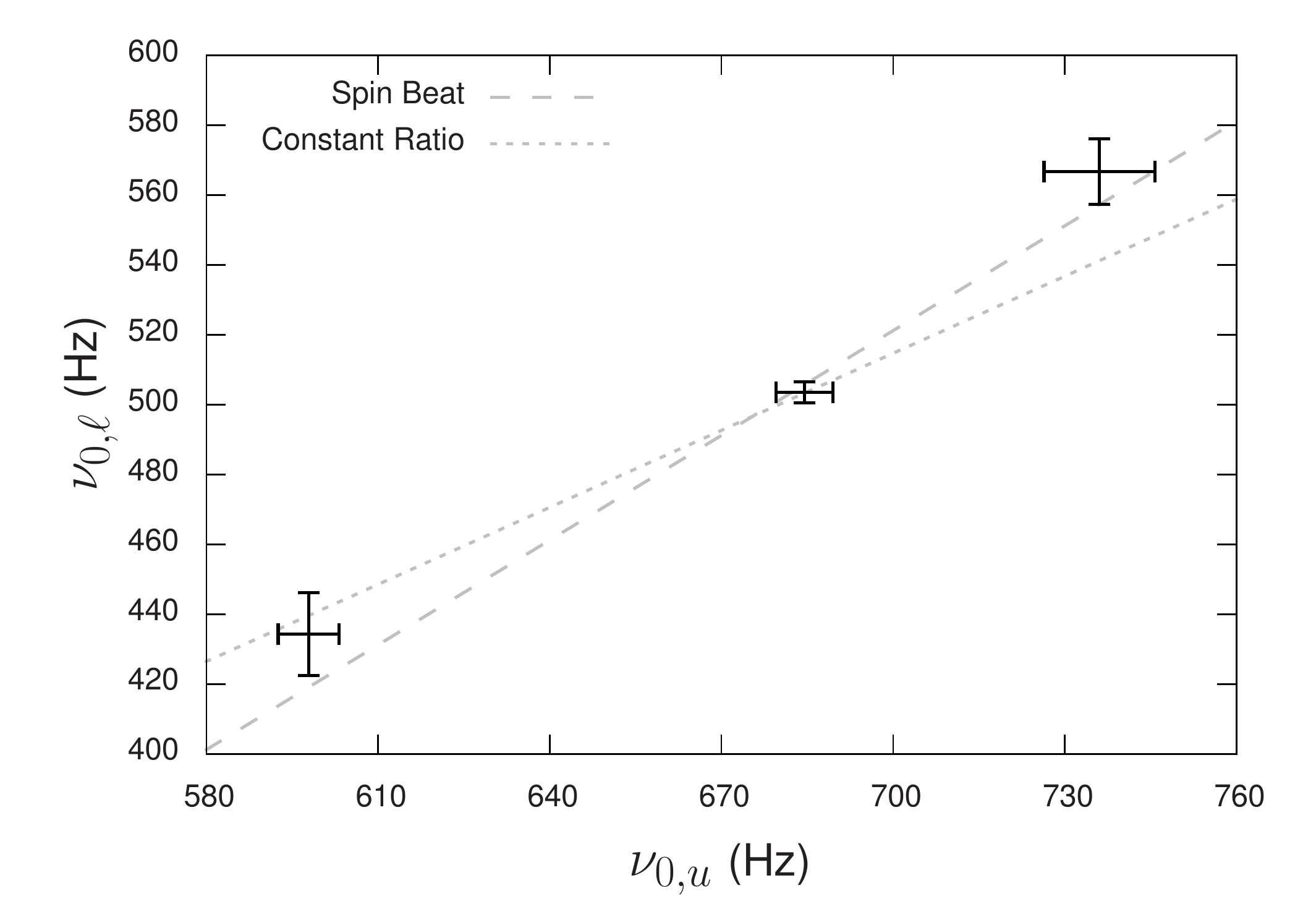}
				\caption{
					Twin kHz QPO centroid frequency trend in SAX~J1808. The dashed line shows the linear
					relation $\nu_\ell = \nu_u - 0.446\nu_{\text{spin}}$, and the dotted line
					gives a proportionality at the constant ratio of 1.36.
				}
				\label{fig:kHzQPO}
			\end{figure}

			Considering the centroid frequencies of all three detections of the lower kHz 
			QPOs we find that the twin kHz QPOs move over a frequency range of about 150~Hz 
			(see Figure~\ref{fig:kHzQPO}). 
			This trend can be fitted with a constant frequency difference of 
			$\nu_{0,u}-\nu_{0,\ell} = (0.446\pm0.009)\nu_{\text{spin}}$, which is
			\textit{inconsistent} with being exactly half the spin frequency, but, being
			slightly smaller, may still be explained with a beat frequency model \citep{Lamb2003}.
			Alternatively the measurements may also be explained by a constant $\nu_{0,l}/\nu_{0,u}$
			ratio of $1.36\pm0.01$, close to, but inconsistent with the proposed 7:5 resonance 
			ratio. We do note that the obtained frequency trend is tentative, as it is based 
			on a small number of measurements and may be subject to systematic uncertainties that 
			are not captured by the quoted errors.
			
		\subsubsection{The 410~Hz QPO}
		\label{sec:410discussion}
            Returning to the frequency relations shown in Figure~\ref{fig:AtollScheme}, we 
            now consider the $\sim$410~Hz QPO. 
            The characteristic frequency of $L_{410}$ correlates with the upper 
            kHz QPO, although this relation is less steep than the others. 
            It was noted by \citet{Wijnands2003} that the difference between the 
            410~Hz QPO frequency and the 401~Hz spin frequency follows a quadratic 
            trend with $\nu_u$. In Figure \ref{fig:AtollScheme} we therefore subtract
            the spin frequency from the measured $L_{410}$ frequency,
            revealing a frequency correlation for the 410~Hz~QPO that is
            similar to that of the other components. This correlation suggests that 
            the 410~Hz QPO is a sideband of the pulsations caused by the beat 
            with a low frequency disk component.

			Because we consider rotation frequencies, a beat signal may appear
			either as lower or upper sideband, depending on the relative
			sense of rotation. Since $L_{410}$ is observed as an upper sideband 
			only, this suggests a beat with a quasi-periodic rotational low frequency disk phenomenon,
			$L_{410-401}$, that is retrograde (counter-rotates) with respect to 
			the neutron star spin.  
						
			We searched for the associated low frequency $L_{410-401}$ component at 9~Hz in the power 
			spectrum, both directly and using a shift-and-add method \citep{Mendez1998}, but did not detect 
			it. Given that $L_{410}$ has a very small amplitude, it is possible that 
			the original $\sim9$~Hz signal is below the detection limit in this
			frequency region, which is dominated by the prominent $L_{b}$ and 
			$L_h$ components. In the following we shall refer to the $L_{410-401}$ component
			keeping in mind that this inferred $\sim9$~Hz counter-rotational disk phenomenon was not,
			in fact, detected as a power spectral component, but seen only as a $\sim410$~Hz 
			beat with the spin.  

			It was noticed by \citet{Straaten2005} that the $L_{410-401}$ frequency
			is about half $\nu_h$. If we fit the relation between these two characteristic
			frequencies we indeed obtain a proportional relation with a slope of $0.53\pm0.03$.
			Using this relation we can predict the frequency at which $L_{410}$ should appear, 
			allowing for a shift-and-add search in the observations where $L_{410}$ is not 
			directly observed.
			
			In Section~\ref{sec:410results} we noted that $L_{410}$ is detected
			for $320\leq\nu\leq401$~Hz if the source flux is greater than $\sim40$~mCrab.
			We therefore first select all the low flux observations with $\nu_u$ in this 
			frequency range for which
			$L_{410}$ was not detected. A shift-and-add search revealed a marginal
			feature ($2.8\sigma$) at $\nu_{\rm spin}+\nu_h/2$, suggesting $L_{410}$
			might also present below $\sim40$~mCrab. The observed feature has a fractional rms of
			$1.4$\% for a fixed $Q=178$. While this quality factor is higher than the typically
			observed $Q\simeq80$, it is consistent within typical errors.
			
			A shift-and-add search for all observations with $\nu_{0,u}\geq\nu_{\rm spin}$ and
			a search for all observations with $\nu_{0,u}\leq320$~Hz did not
			produce a detection ($<1\sigma$), giving 95\% upper limits on the
			fractional rms of less than 1\%. We note that this could either mean that
			the $L_{410}$ feature is not present in those observations, is too weak to be detected, 
			or that the proposed relation with $\nu_h$ is not correct. 
			
		\subsubsection{The LF QPO}
			The narrow QPO at $\sim50$~Hz, which we labeled $L_{\LF}$, was identified
			as such by \citet{Straaten2005} based on extrapolating the $L_{\LF}$ versus 
			$L_h$ relation found in low luminosity bursters (1E~1724--3045, 
            GS~1826--24). It was shown by \citet{Altamirano2008} that the index
            of this power-law is consistent with 1, implying the relation is 
            consistent with a simple proportionality. Extending this dataset
            with the additional measurements from our work we find that a proportional
            function performs well and that the proportionality ratio is 
            $0.65\pm0.02$. 

\begin{figure}[t]
				\includegraphics[width=\linewidth]{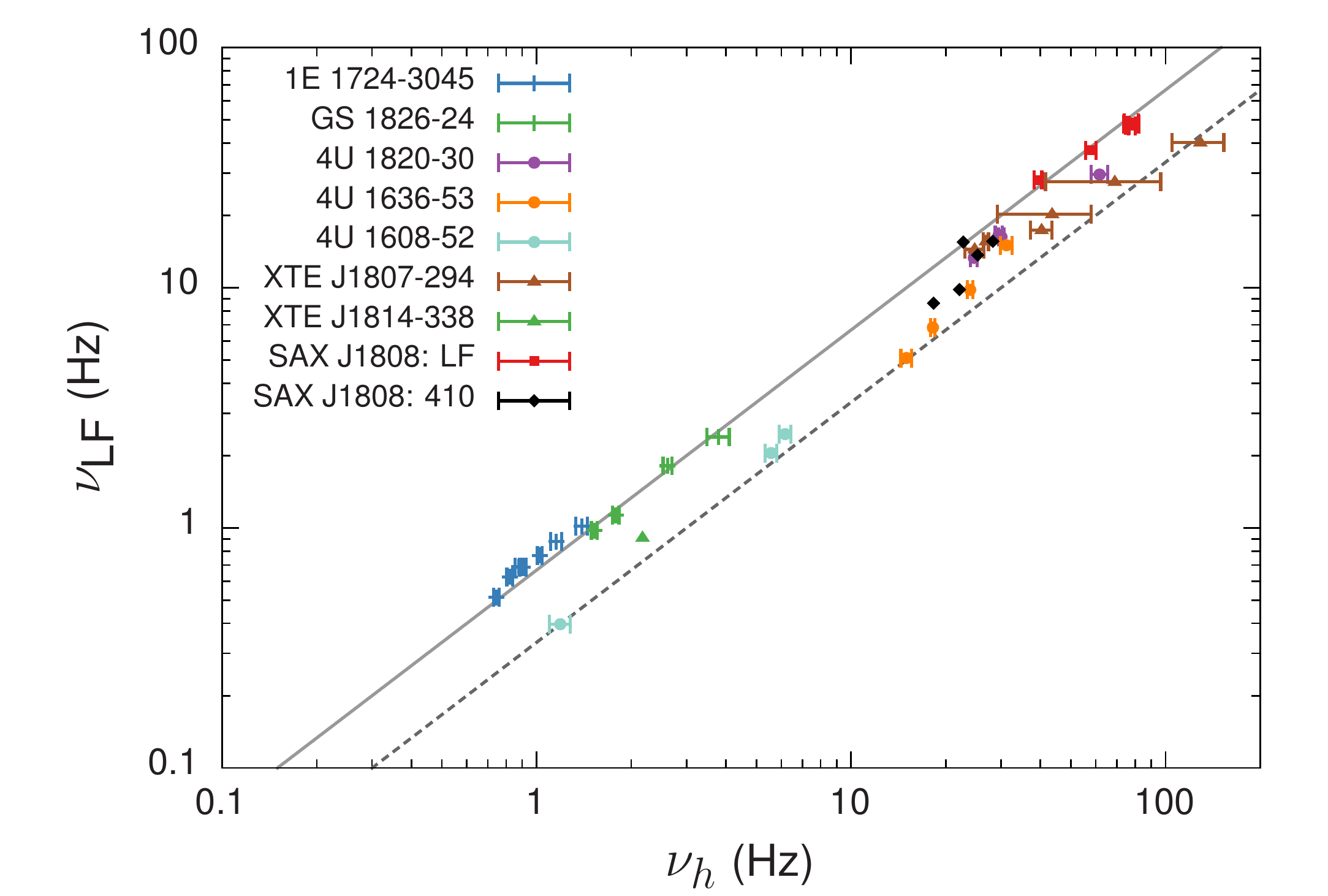}
				\caption{
					Characteristic frequency of $L_{\LF}$ versus $L_h$ for SAX~J1808 and
					several other atoll sources. The solid line show a proportionality ratio of 0.66 
					and the dashed line shows a proportionality ratio of 0.33. \\
				}
				\label{fig:NarrowQPO}
			\end{figure}

            While the proportionality is consistent with 2/3, this small integer ratio of $L_{\LF}$ 
            with $L_h$ is only seen in SAX~J1808, 1E~1724--3045, and 
            GS~1826--24 (see Figure~\ref{fig:NarrowQPO}) and only when considering
            characteristic frequencies ($\nu_{max}$). 
            Relativistic resonance models \citep{Kluzniak2001}, which use centroid 
            frequencies, are therefore not applicable.
            
            While other atoll sources do not show a LF QPO that follows the same relation with $L_h$
            we see in SAX~J1808, it has been suggested that some show a sub-harmonic ($L_{\LF/2}$,
            proportionality ratio of $0.33$)
            rather than $L_{\LF}$ \citep{Straaten2005}. However, many other sources have shown
            similar QPOs \citep[e.g.][]{Altamirano2005, Altamirano2008} with frequencies drifting smoothly 
            between $L_{LF}$ and $L_{LF/2}$, crossing the frequency range of the $L_{410-401}$ 
            component (Figure~\ref{fig:NarrowQPO}).
            This suggests $L_{\LF}$, $L_{\LF/2}$, and $L_{410-401}$ might be the same component.

		\subsubsection{LF and 410~Hz QPO}
			If $L_{\LF}$ and $L_{410-401}$ are the same feature then
			this raises the question why the LF component is seen directly in the power spectrum
			and $L_{410-401}$ only as a sideband of the pulsations and why $L_{\LF}$ and 
			$L_{410}$ are never seen simultaneously. 
			
			The 410~Hz QPO is only observed when the upper kHz QPO centroid frequency is below
			the spin frequency. We recently found that when the upper kHz QPO frequency moves 
			above the spin frequency the pulse amplitude decreases by a factor of $\sim2$ 
			\citep{Bult2015}. If this drop in pulse amplitude also applies to the amplitude 
			of the beat signal, then the power of the 410~Hz QPO would drop by a factor of $\sim4$,
			putting it below the detection threshold of those observations where the upper kHz QPO 
			frequency is above the spin frequency. 
			
			As the upper kHz QPO increases further, and presumably the
			inner edge of the disk moves closer to the neutron star surface, neither the LF QPO nor
			the 410~Hz feature are observed. Finally, when source transitions to the the lower-left banana, 
			for the highest $L_u$ frequencies observed in SAX~J1808 and thus when the accretion disk has probably
			advanced closest to the neutron star surface, the LF QPO appears. If the LF QPO 
			originates from the inner region of the accretion disk, for instance through precession, 
			then the increased disk temperature may be the reason that the LF QPO is observed directly, 
			while the pulsations are still too weak to power a detectable beat signal. 
			
	\subsection{Precession Models}
        A popular interpretation of the frequency relations associates QPO
        frequencies with relativistic orbital and precession frequencies
        \citep{Stella1998,Stella1999}. In this model the LF QPO frequency corresponds to the nodal
        precession frequency (or its second harmonic) of a test mass orbiting at the disk 
        inner radius as caused by relativistic frame dragging, which in
        the weak field approximation is known as Lense-Thirring (LT) precession. 
        More recent versions of this model consider the LF QPO to be caused by a geometrically 
        thick inner flow which, also due to frame dragging, precesses as a solid body 
        \citep{Fragile2007,Ingram2009}, and produces the correct frequencies in black hole and
        neutron star systems \citep{Ingram2009,Ingram2010}, while naturally
        explaining a range of QPO properties \citep{Ingram2012}.
		
        For atoll sources specifically the relativistic precession models attribute
        $L_{\LF}$ to LT precession. If $L_{LF}$ and $L_{410-401}$ are the same
        component, then the interpretation that the 410~Hz QPO arises due to a retrograde beat with
        the neutron star spin, implies that the precession too must be retrograde. However,
        LT precession is always prograde with the stellar rotation \citep{Merloni1999,
        Markovic2000}, excluding frame dragging as the main cause for the precession.
        
        For neutron stars, however, frame dragging is not the only cause of precession.
        The neutron star oblateness and its magnetic field may also apply torques to the inner
        accretion disk, which cause a retrograde (negative) contribution to the precession
        frequency \citep{Lai1999, Shirakawa2002}. 
        In the magnetic precession model the stellar magnetic field induces a warp in 
        a geometrically thin accretion disk, which in turn undergoes retrograde precession. 
        However, with a power-law index of $\sim4.6$ \citep{Shirakawa2002}, pure magnetic 
        precession predicts a much steeper relation with the upper kHz QPO than we observe. 
        Classical precession (due to the oblateness), on the other hand, predicts a power-law 
        index of $\sim2.3$, which is closer to what we observe.
		It would be interesting to investigate if in neutron star systems the combination of
		relativistic, classical and magnetic torque contributions can 
		produce a retrograde variant of the LT thick flow precession seen by \citet{Fragile2007} 
		in simulations of black hole disks, with all the attractive properties gained from it.

\section{Conclusions}
    We have presented a complete overview of the aperiodic variability of SAX~J1808 
    as observed with \textit{RXTE} over the course of 14~years. We find that SAX~J1808
    is mainly observed as an atoll source in the island state, and sometimes
    transits to the lower-left banana or the extreme island state. The
    individual outbursts are very similar in terms of their light curve and 
    color evolution and, aside from the 1--5~Hz high luminosity flaring observed
    in 2008 and 2011, show a similar evolution of the power spectra as well.
	
    Considering the power spectra of all outbursts we find that 
    all characteristic frequencies are correlated with $\nu_{u}$, and that
    the correlations show a break when the source state changes between
    the island and extreme island states around $\nu_u\simeq250$~Hz. This break 
    disappears when considering the relations with $\nu_b$ instead, suggesting that
    it is $\nu_u$ that shows a state dependent relation with the other variability 
    components.
    
    We find that at $\nu_{u}\gtrsim 550$~Hz the frequency of the hHz component in SAX~J1808 shows an 
    unexpected correlation with $\nu_u$, that roughly follows the frequency relations
    of the other components. We suggest that this correlation could be due to a blend
    with an unresolved $\low$ component or that it could be intrinsic to the hHz component
    itself.    
    We also considered the relation between extreme island state $\low$ component and the 
    lower kHz QPO, but found no definitive evidence in favor or against the two being the
    same. 
    
    We also presented new measurements of the lower kHz QPO, which, for the first time,
    allows us to probe the frequency evolution of the twin kHz QPOs in this source. Our
    observations are consistent with a twin kHz QPO frequency separation of 
    $(0.446\pm0.009)\nu_{\text{spin}}$, which is inconsistent with half the spin frequency,
    but tentatively so, as the fit is based on a small number of measurements.
    
    Additionally, we presented the first detailed study of the 410~Hz QPO and found that it
    appears only when $320 \mbox{\ Hz}\leq\nu_{u}\leq\nu_{\rm spin}$. 
    Subtracting the spin frequency from the measured $\nu_{410}$ we find that the
    resulting frequency matches the overall correlation trends, which strongly suggests that the
    410~Hz QPO is caused by a retrograde beat against the 401~Hz spin
    frequency, even though the original $\sim9$~Hz signal is never seen. 
	
    Comparing the measured hump, LF and 410~Hz beat QPO with $L_{\LF}$
    detections in other neutron star systems we suggest that the LF and 410~Hz beat 
    QPOs might be the same component. We suggest this QPO might be related to retrograde 
    nodal precession caused by the (retrograde) classical and magnetic precession
    dominating over the (prograde) LT precession.

\acknowledgments
The authors acknowledge support from the Netherlands Organisation for Scientific Research (NWO).

\bibliographystyle{apj}

\end{document}